%% file: main.tex
\documentclass{article}

\usepackage{PRIMEarxiv}

\usepackage[utf8]{inputenc} 
\usepackage[T1]{fontenc}    
\usepackage{hyperref}       
\usepackage{url}            
\usepackage{booktabs}       
\usepackage{amsfonts}       
\usepackage{nicefrac}       
\usepackage{microtype}      
\usepackage{lipsum}
\usepackage{fancyhdr}       
\usepackage{graphicx}       
\usepackage{rotating}       
\usepackage{threeparttable} 
\usepackage{array}          
\graphicspath{{figures/}}     
\usepackage{multirow}
\usepackage{subcaption}
\usepackage{longtable}
\usepackage[title,titletoc]{appendix}
\usepackage{amsmath}
\usepackage{pdflscape}
\usepackage{geometry}

\usepackage{changes}


\title{Modeling Drug Overdose Mortality Levels Using Social Vulnerability Indicators in U.S. Counties: A Double Machine Learning Approach for Causal Analysis 
}

\author{
  Qunhui Guo, Jiongbai Liu \\
  Decision Analytics Lab\\
  Northeastern University \\
  Boston\\
  \texttt{guo.qun@northeastern.edu} \\
  \texttt{liu.jio@northeastern.edu} \\
   \And
  Md. Noor-E-Alam\\
  Decision Analytics Lab\\
  Northeastern University \\
  Boston\\
  \texttt{md.alam@northeastern.edu} \\
}

\begin{document}
\maketitle

\begin{abstract}
Drug poisoning mortality (DPM) remains a leading cause of death among working-age Americans, and its burden falls unevenly across communities. Existing county-level studies of social vulnerability and overdose have largely reported predictive or associational relationships, while the shape of the adjusted exposure-response relationship between overall social vulnerability and mortality remains unclear.
We examined the CDC/ATSDR Social Vulnerability Index (SVI) and county drug poisoning mortality across three two-year windows spanning 2016–2021. 
The associational analysis compared linear regression, Random Forest, and XGBoost models of log mortality using period-specific SVI subfeatures and county covariates.
The selected XGBoost models were interpreted with SHAP values. 
The DML analysis estimated the exposure-response relationship between SVI and DPM and tested the linearity restriction. 
The socioeconomic theme vulnerability contributed positively to predicted mortality, with disability and unemployment among the strongest subfeatures. While racial and ethnic minority composition contributed negatively in all three periods.
Summing the four theme contributions reduced the overall-index attribution by 34\% to 36\%, indicating substantial offsetting. 
Linearity was rejected in every window ($p<10^{-26}$). The adjusted relationship increased through the middle of the SVI distribution, and declined in the upper tail. This pattern remained stable both before and after the outbreak of the epidemic.
Overall SVI therefore captured meaningful variation in overdose mortality but did not function as a uniformly increasing risk scale across its full range. 
Component-level interpretation and current overdose surveillance provide important context for interpreting the composite index.

\end{abstract}

\keywords{Causal Analysis \and Double Machine Learning \and Public Health}

\input{Introduction}
\input{Methods}

\input{Results}
\input{Discussion}

\input{Conclusion}

\bibliographystyle{unsrt}  
\bibliography{bibliography}  

\clearpage

\begin{appendices}
\input{Appendix}
\end{appendices}

\end{document}

%% file: Introduction.tex
\section{Introduction}

Drug overdose is among the most consequential public health problems in the United States. 
The age-adjusted drug overdose death rate increased from 6.1 deaths per 100,000 population in 1999 to a peak of 32.6 in 2022 \cite{warner2014trends, spencer_garnett_minino_2023_drug_overdose}. The epidemic has evolved in successive waves: prescription opioids in the early 2000s, heroin in the early 2010s, and most recently illicitly manufactured fentanyl, often combined with stimulants such as methamphetamine and cocaine \cite{spencer_garnett_minino_2023_drug_overdose, nida_overdose_death_rates_2024}. 
After more than two decades of escalation, the national trend began to reverse, with the rate declining to 31.3 in 2023 and 23.1 in 2024 \cite{garnett_drug_overdose_2023_2024}. Despite this recent decline, overdose mortality remains the leading cause of death among Americans aged 18 to 44 \cite{cdc_overdose_decline_2025}.

These national trends, however, mask substantial geographic variation in overdose mortality across U.S. counties.
County-level clusters of elevated unintentional drug poisoning mortality (DPM) proliferated across the western United States from 2013, with further clusters emerging in the east after 2016 \cite{escobar2025spatiotemporal}. Other small-area analyses have also identified persistent county-level hot spots in drug poisoning mortality \cite{rossen2014hot}. 
The geographic distribution of overdose mortality also varies along the urban-rural continuum, with socioeconomic distress playing a larger role in rural counties and opioid supplying a larger role in urban counties \cite{mack2017illicit, monnat2019contributions}. Much of this heterogeneity occurs below the state level, and county social vulnerability tracks substantial differences in overdose deaths and treatment admissions \cite{cooper2025substance}. A systematic review of spatial studies finds social and economic conditions consistently associated with overdose mortality across small areas \cite{fink2024social}; County characteristics explain much of the geographic difference in drug-related mortality \cite{monnat2018factors}. These findings support the county as a meaningful unit of analysis and point to the importance of community-level social conditions in shaping geographic differences in overdose mortality.

Understanding these geographic differences requires considering factors operating at both individual and community levels. Opioids account for the majority of drug overdose deaths in the United States, making opioid-specific evidence particularly relevant to the broader overdose burden. At the individual level, lower income, lower educational attainment, incarceration, housing and employment instability, and limited social support are each associated with elevated overdose risks \cite{van2020socioeconomic}. These individual-level risks are embedded within broader community contexts. Neighborhood economic disadvantage, the built environment, and the local availability of health and social resources are independently associated with drug poisoning mortality after individual factors are accounted for \cite{flores2020associations, sistani2023social}. Counties with greater social capital also experience lower overdose mortality \cite{zoorob2017bowling, verinumbe2025association}. 
The Social Vulnerability Index (SVI), developed by the Centers for Disease Control and Prevention's Agency for Toxic Substances and Disease Registry (CDC/ATSDR), summarizes county vulnerability using census-based indicators grouped into four themes: socioeconomic status, household characteristics, racial and ethnic minority status, and housing type and transportation. It assigns each county percentile ranks ranging from 0 to 1 for overall vulnerability and for its component measures\cite{cdc2020svi}, with higher values indicating greater social vulnerability. Originally developed to identify communities requiring support during disasters, the SVI has since been applied more broadly as a measure of place-based health risk \cite{khazanchi2020county, khan2021social}. 
It is publicly maintained, nationally standardized, and already incorporated into federal and state resource-allocation decisions. Consequently, researchers have increasingly applied the SVI to examine overdose-related outcomes \cite{stokes2023county, tiruwa2026social}.

This growing body of work broadly falls into two lines of research.
The first examines associations between social vulnerability and overdose burden across diverse settings. County-level SVI has been consistently linked to fatal and nonfatal overdose outcomes, spanning drug poisoning mortality, cocaine-specific mortality, nonfatal overdose emergency department visits, and national disparities in drug and alcohol overdose deaths \cite{tatar2023social, sung2025effect, grobman2026social, cooper2025substance}.
Related supply-side work further reveals higher opioid dispensing rates in more vulnerable counties, showing that community vulnerability compounds local prescription opioid availability to elevate individual risk \cite{bounthavong2025association, el2023association}. 
Interpretation of SVI-based findings also requires attention to the changing demographic geography of the overdose epidemic during this period. Black overdose mortality remained relatively stable through much of the 2000s, increased sharply after 2017, and passed the White rate for the first time in 2019 \cite{smith2025widening}. The demographic composition of overdose mortality therefore shifted within the study period.
The second develops machine-learning models to predict county- or neighborhood-level overdose mortality, and ranks structural predictors with feature-importance or SHAP methods \cite{tatar2023social, kumar2025opioid, schell2022identifying, ebrahimi2025spatial, DEAS2026101759}. 
Across these studies, disability repeatedly emerges as a leading predictor. 

Despite these advances, three key gaps remain in the literature.
First, existing work has not fully characterized how the hierarchical structure of the SVI interacts with the direction of component-level associations with overdose mortality. Overall SVI combines four themes whose relationships with overdose mortality may differ in both magnitude and direction. Prior machine-learning studies have examined individual SVI components, but their interpretation has largely focused on identifying influential predictors. Tatar et al. ranked SVI features using XGBoost gain \cite{tatar2023social}, while Deas et al. similarly examined individual SVI characteristics using XGBoost gain and a Shapley-based feature attribution method \cite{DEAS2026101759}. Deas et al. additionally examined directional patterns by comparing SVI characteristics across counties with anomalously high and low mortality, but did not evaluate how signed model contributions combine across the SVI hierarchy.
Other county and small-area studies have recovered direction, through regularized-regression signs or SHAP values applied to general county features \cite{schell2022identifying, kumar2025opioid}. However, prior studies have not systematically aggregated signed component-level contributions to themes and subsequently to the overall SVI within a single predictive model. Without this hierarchical view, an aggregate association carries no information about whether the components reinforce or offset one another, leaving a weak composite association indistinguishable from strong, opposing components that cancel.

Second, existing studies remain explicitly predictive or associational, the causal exposure-response relationship between county social vulnerability and drug poisoning mortality has not been formally estimated\cite{ ebrahimi2025spatial, tiruwa2026social}.
Prior county studies have examined overdose mortality in relation to regional patterns, socioeconomic conditions, segregation, and SVI-based vulnerability measures \cite{monnat2018factors, segel2024understanding, frankenfeld2019county, tatar2023social, sung2025effect, grobman2026social}. 
However, the major SVI studies characterize these relationships as predictive or associational, and several explicitly note that their observational or cross-sectional designs do not support causal interpretation \cite{tatar2023social, sung2025effect}. To our knowledge, no prior study has estimated how drug poisoning mortality changes across levels of overall SVI under an explicit adjustment set and stated causal identification assumptions.

Existing evidence is further constrained by rigid functional-form assumptions. Prior analyses have typically operationalized overall SVI using vulnerability categories or a single linear term. Categorization yields only a coarse representation across the vulnerability spectrum, whereas a linear specification imposes a constant slope that cannot detect potentially complex, nonlinear patterns.
Double machine learning (DML) provides a framework for causal estimation under selection on measured covariates while allowing flexible machine-learning estimation of the nuisance functions and reducing sensitivity of the target estimate to regularization and overfitting bias \cite{chernozhukov2018double}. While continuous-exposure extensions of DML have recently been developed \cite{klosin2021automatic, colangelo2026double}, they have not yet been applied to estimate the adjusted SVI-mortality exposure-response relationship.

Third, the temporal scope has been limited. Studies of social vulnerability and overdose during the COVID-19 pandemic have mostly been confined to single cities or localized areas \cite{knoebel2023impact, barboza2022variability}. National work on nonfatal overdose has acknowledged leaving the variation in the 2020 SVI-overdose association undescribed, noting that the pandemic may itself have reshaped community vulnerability \cite{stokes2023county}. Whether the adjusted relationship persists across distinct, vintage-aligned pre-pandemic and pandemic periods at the national scale therefore remains an open question.

This study addresses these gaps through a two-part analytical framework applied to three vintage-aligned two-year windows spanning 2016 to 2021. 
The first part characterizes the association between social vulnerability and drug poisoning mortality (DPM). Using machine learning models interpreted with SHAP, we examined the magnitude and direction of contribution from SVI subfeatures and aggregated these contributions to the theme and overall-index levels.
The second part addresses the causal question by estimating the exposure-response relationship between overall SVI and DPM using partialling-out double machine learning under explicit identifying assumptions. We retain SVI as a continuous exposure and formally test whether the exposure–response relationship departs from linearity.
Because the SVI is a within-vintage percentile ranking, the estimand is a contrast between counties at different positions in the vulnerability distribution rather than the effect of an intervention that changes a county's score.
Applying both analyses across the three windows further allows us to assess whether the relationship persists across the pre-pandemic fentanyl era and the onset of the COVID-19 pandemic.

%% file: Methods.tex
\section{Methods} 
\subsection{Study Data}
This study used a county-level observational design to examine both the association and the causal relationship between social vulnerability and DPM in the United States.
Analyses were conducted separately for three two-year periods: 2016-2017, 2018-2019, and 2020-2021. Each period was analyzed as a separate cross-sectional dataset. This period-specific design allowed comparisons across two pre-pandemic windows and the early COVID-19 pandemic period without assuming a constant vulnerability–mortality relationship across these distinct epidemiologic contexts.

Social vulnerability was measured using the CDC/ATSDR Social Vulnerability Index \cite{cdc2020svi}. The SVI summarizes county-level vulnerability using census-based social and demographic indicators and reports percentile rankings for overall vulnerability, four vulnerability themes, and individual component measures. The 2016, 2018, and 2020 SVI releases were aligned with the 2016–2017, 2018–2019, and 2020–2021 mortality windows, respectively. Overall SVI was used as the primary measure of county social vulnerability, while theme- and component-level measures were used to characterize which dimensions of vulnerability were most strongly associated with DPM.

Drug poisoning mortality (DPM) data were obtained from the National Center for Health Statistics (NCHS) county-level drug poisoning mortality dataset \cite{rossen2022county}, which is derived from National Vital Statistics System (NVSS) multiple cause-of-death mortality records. Drug poisoning deaths are identified using ICD-10 underlying cause-of-death codes X40-X44, X60-X64, X85, and Y10-Y14, covering unintentional, suicide, homicide, and undetermined-intent drug poisonings. To improve the stability of estimates in counties with sparse annual death counts, NCHS provided model-based county estimates generated using hierarchical Bayesian models with spatial and temporal random effects. To construct the two-year analytic periods, annual mortality rates were combined within each two-year period using population-weighted averages. The dataset also provides urban-rural classification and census division, which were retained as geographic covariates in the analysis. 

Additional county-level covariates captured demographic, geographic, and structural factors. Resident population, daytime population, and land area were obtained from the corresponding CDC/ATSDR SVI releases, from which population density and the daytime-to-resident population ratio were derived. Because these metrics were sourced from the SVI datasets, their vintages naturally aligned with the exposure.
Sex measures were obtained from American Community Survey tables \cite{census2021acsS0101}, and sex ratio was calculated after summing male and female population counts across the two years. County-level structural characteristics were obtained from the U.S. Department of Agriculture (USDA) Economic Research Service (ERS) County Typology Codes, 2015 Edition, which classify counties by primary economic dependence, including farming, mining, manufacturing, government, and recreation, and identify demographic characteristics including population loss and retirement-destination status \cite{usda_ers_typology_2015}. Because the ERS typology codes are based on a single 2015 release, these values were treated as time-invariant across the three analytic windows.

Datasets were merged by county FIPS code within each analytic period. Counties were excluded if they lacked DPM, required SVI measures, or covariates required for the primary adjusted analysis. After data linkage, exclusions, and preprocessing, the final analytic samples included 3,130 counties in 2016-2017, 3,129 counties in 2018-2019, and 3,130 counties in 2020-2021. The small differences in sample composition involved three counties. Autauga County, Alabama, was excluded from the 2016–2017 and 2018–2019 analyses because at least one required measure was unavailable, but was included in 2020–2021. Rio Arriba County, New Mexico, was included in 2016–2017 and 2020–2021 but excluded from 2018–2019 because of missing required data. Valdez-Cordova Census Area, Alaska, was included in the two earlier windows but not in 2020–2021 following its dissolution and replacement by the Chugach and Copper River census areas \cite{uscensus_county_changes}, for which comparable NCHS estimates were unavailable. Across the three period-specific samples, 3,128 counties were observed in all three windows. Because each window was analyzed independently as a cross-sectional dataset, all eligible counties within each period were retained.

\subsection{Measures}
\label{sec: measures}
The primary outcome was county-level DPM, defined as the population-weighted average of the annual mortality rates within each two-year analytic window. DPM was log-transformed for the primary analyses to reduce residual skewness and stabilize residual variance across fitted values.

The primary exposure was the overall SVI. 
The associational analysis additionally examined the period-specific SVI subfeatures to characterize the contribution of individual dimensions of vulnerability to predicted DPM. The composition of these subfeatures changed in the 2020 release, which replaced the poverty measure with the percentage of persons below 150\% of the federal poverty level, removed per capita income, added housing cost burden, incorporated uninsured status into the socioeconomic theme, and moved limited English proficiency from the minority-status theme to the household-characteristics theme.
In the Double Machine Learning analysis, overall SVI was treated as a continuous exposure.

The adjustment set comprised county-level demographic, structural, and geographic characteristics selected as potential confounders of the relationship between SVI and DPM.
These included resident population, population density, the daytime-to-resident population ratio, sex ratio, indicators for population-loss and retirement-destination counties, and three categorical variables: USDA ERS county economic type (reference: Nonspecialized), urban–rural classification (reference: Noncore), and census division (reference: New England). Because resident population and population density were highly right-skewed, both were log-transformed before modeling to reduce the influence of extreme values. The same transformed definitions were used across the regression, tree-based learners, and Double Machine Learning analyses to maintain a consistent covariate specification. Collectively, these variables captured differences in population composition, settlement patterns, economic structure, and broad geographic context across counties. The same non-SVI adjustment set was used in both analyses.

Detailed definitions of the measures are provided in Appendix Table \ref{tab:variable-definitions}.

\subsection{Analytic Methods}
\label{sec:Analytic Methods}
The analytic framework comprised two complementary components. The associational analysis examined the association between SVI and DPM, and the causal analysis estimated the causal effect of overall SVI on DPM using DML.

For the associational analysis, the complete-case sample in each analytic window was partitioned $70/30$ into training and test sets. Three models were fit to the training set with the period-appropriate SVI subfeatures and the covariate set as predictors of $\log\text{DPM}$: linear regression, XGBoost and Random Forest. The regression model was  fit directly. Hyperparameters for XGBoost and Random Forest were tuned using 5-fold cross-validated randomized search within the training set. Model performance was evaluated on the test set using $R^2$, root mean squared error (RMSE), and mean absolute error (MAE). RMSE was prespecified as the primary model-selection criterion, with $R^2$ and MAE reported as complementary measures of predictive performance. The model with the lowest RMSE was selected for interpretation.

Tree-based model interpretation used SHAP values computed on the test set \cite{lundberg2017unified}. We ranked feature importance using the mean absolute SHAP values across counties in the test set. 
Exploiting the additive property of SHAP, we summarized contributions at higher levels of the SVI hierarchy by summing signed SHAP values within each county across the subfeatures of each theme and, and across all subfeatures for the overall index.
Direction was summarized using the Spearman correlation $\rho$ between each subfeature, theme, or overall SVI ranking and its corresponding SHAP contribution, a strategy previously used to characterize the direction of SHAP attributions \cite{joachim2026evaluating}. 
Positive correlations indicated that higher values contributed to higher predicted DPM, whereas negative correlations indicated the opposite. 
Values near zero indicated no consistent monotonic direction and did not imply low predictive importance.
For binary indicators, direction was summerized using mean SHAP contributions for the indicator category relative to its reference category. SHAP values described the fitted predictive model and were not interpreted as causal effects.

To align the associational analysis with the exposure used in the DML analysis, we fitted an additional XGBoost model using overall SVI and the same non-SVI covariates to predict log DPM. The same train/test assignments and XGBoost tuning procedure were used as in the primary associational analysis.

Additionally, two sensitivity analyses were applied to the association analysis across all three windows. The first restricts the predictor set to the SVI subfeatures alone by excluding non-SVI covariates. The second replaces log DPM with the untransformed DPM while retaining the full predictor set.


The causal analysis examined the effect of overall SVI on log DPM using a partialling-out Double Machine Learning estimator. The treatment was overall SVI and the outcome was log DPM. The adjustment set $X$ comprised the non-SVI county covariates defined in Section ~\ref{sec: measures}, selected as plausible common causes of overall SVI and DPM.
Because overall SVI is a relative percentile ranking constructed from multiple component measures, the estimand was interpreted as a contrast in expected log DPM between counties at different positions in the overall SVI distribution after adjustment for $X$, rather than as the effect of changing a county's SVI itself.
SVI theme rankings and subfeatures were excluded from $X$ because they are constituent measures used to construct overall SVI; conditioning on them would adjust for components of the treatment rather than for confounders.

For the linear partially linear specification, $Y$ denoted log DPM, $D$ denoted overall SVI, and $X$ denoted the adjustment set.
Two nuisance functions were estimated using 5-fold cross-fitting: an outcome model $E[Y|X]$ and a treatment model $E[D|X]$. For each nuisance function, Elastic Net, XGBoost, and Random Forest were evaluated using cross-fitted RMSE over the full analytic sample, and the learner with the lowest RMSE was selected separately for the outcome and treatment models. The selected predictions were used to residualize $Y$ and $D$ for the final DML estimation.

The residualized outcome and treatment were used to solve the Neyman-orthogonal partialling-out estimating equation of Chernozhukov et al. \cite{chernozhukov2018double}, yielding $\hat{\theta}$, the estimated linear DML coefficient for overall SVI. Neyman orthogonality makes the estimating equation locally insensitive to first-order errors in the nuisance-function estimates.
Standard errors were calculated using an influence-function-based variance estimator, and 95\% confidence intervals were constructed using a normal approximation. 
For interpretation, estimates were reported for a 0.10-unit increase in SVI, corresponding to 10 percentile points on the 0–1 scale. The associated percent change in the conditional geometric mean of DPM was calculated as $100 \times (\exp(0.10 \hat\theta) - 1)$, with confidence limits transformed accordingly.

The standard partially linear specification assumes a constant slope for overall SVI across the exposure range. We therefore treated $\hat{\theta}$ as a linear benchmark and then evaluated departures from linearity using a basis-expanded partially linear model:
$$Y=h(D)+g(X)+\varepsilon, \qquad h(D)=\sum_{k=1}^{K}\beta_k b_k(D),$$

where $h(D)$ represents the exposure–response function, $g(X)$ represents the covariate component, $b_k(D)$ denotes the $k$th restricted cubic spline basis function with coefficient $\beta_k$, and $\varepsilon$ is the error term.
Here, $K$ denotes the number of spline basis functions. The primary four-knot specification contained one linear and two nonlinear terms. The nonlinear specification used XGBoost for the nuisance regressions under the cross-fitting procedure described above. 
The outcome and each spline basis term were separately predicted from $X$, and their cross-fitted residuals were used to estimate the spline coefficients jointly. Knots were placed at the 5th, 35th, 65th, and 95th percentiles of the SVI distribution within each analytic sample \cite{harrell2015regression}. Departure from linearity was assessed using a joint Wald test of the nonlinear spline terms. The fitted curve was displayed relative to the median SVI within the first quartile of each analytic window.

To facilitate interpretation of the nonlinear relationship, we also estimated partialling-out models in which overall SVI was represented by quartile and quintile indicators, with the lowest category as the reference. Quartiles summarized the broad nonlinear pattern, while quintiles provided finer resolution across exposure levels. 
Because the category coefficients were estimated jointly, adjacent-category contrasts, including Q3 versus Q4 and Q4 versus Q5, were evaluated using their joint covariance matrix rather than inferred from overlap of separate confidence intervals.

A causal interpretation of these estimates requires conditional exchangeability given $X$, positivity, and consistency. DML addresses estimation under these assumptions but does not establish them. 
Empirical overlap for the continuous exposure was evaluated using two complementary diagnostics. First, residual exposure variation was quantified by the following standard-deviation ratio:
$$\mathrm{SD}\{D-\hat E[D|X]\}/\mathrm{SD}(D).$$
The numerator is the standard deviation of residualized SVI, and the denominator is the standard deviation of observed SVI.
Larger values indicate that more variation in SVI remains after conditioning on $X$. Values closer to zero reflect limited residual exposure variation after conditioning on covariates, which would undermine empirical support for estimating the SVI–DPM relationship.
Second, the distribution of observed SVI was evaluated across deciles of predicted SVI, summarizing the median and 5th–95th percentile ranges within each stratum.
These diagnostics were used to assess whether substantial SVI variation remained after conditioning on the adjustment set.

Sensitivity analyses focused on the primary nonlinear specification. 
We examined three- and five-knot spline specifications as alternatives to the primary four-knot model and replaced census division with state indicators to assess sensitivity to finer geographic adjustment. We also re-estimated the model using Elastic Net and Random Forest for the nuisance regressions and repeated the analysis with untransformed DPM. Across these specifications, we compared the test of nonlinearity, the fitted exposure–response curve, the spline contrast between SVI values of 0.25 and 0.75, the Q5-versus-Q4 contrast, and the per-0.10-unit linear benchmark..


%% file: Results.tex
\section{Results}
\subsection{Descriptive Statistics}
Counties were classified into period-specific quartiles of the overall Social Vulnerability Index within each analytic window. Appendix Table \ref{tab:table1_svi_combined} presents county characteristics by quartiles and analysis windows, including components, the outcome, and non-SVI covariates.

\begin{figure}[ht]
\centering
\includegraphics[width=0.6\textwidth]{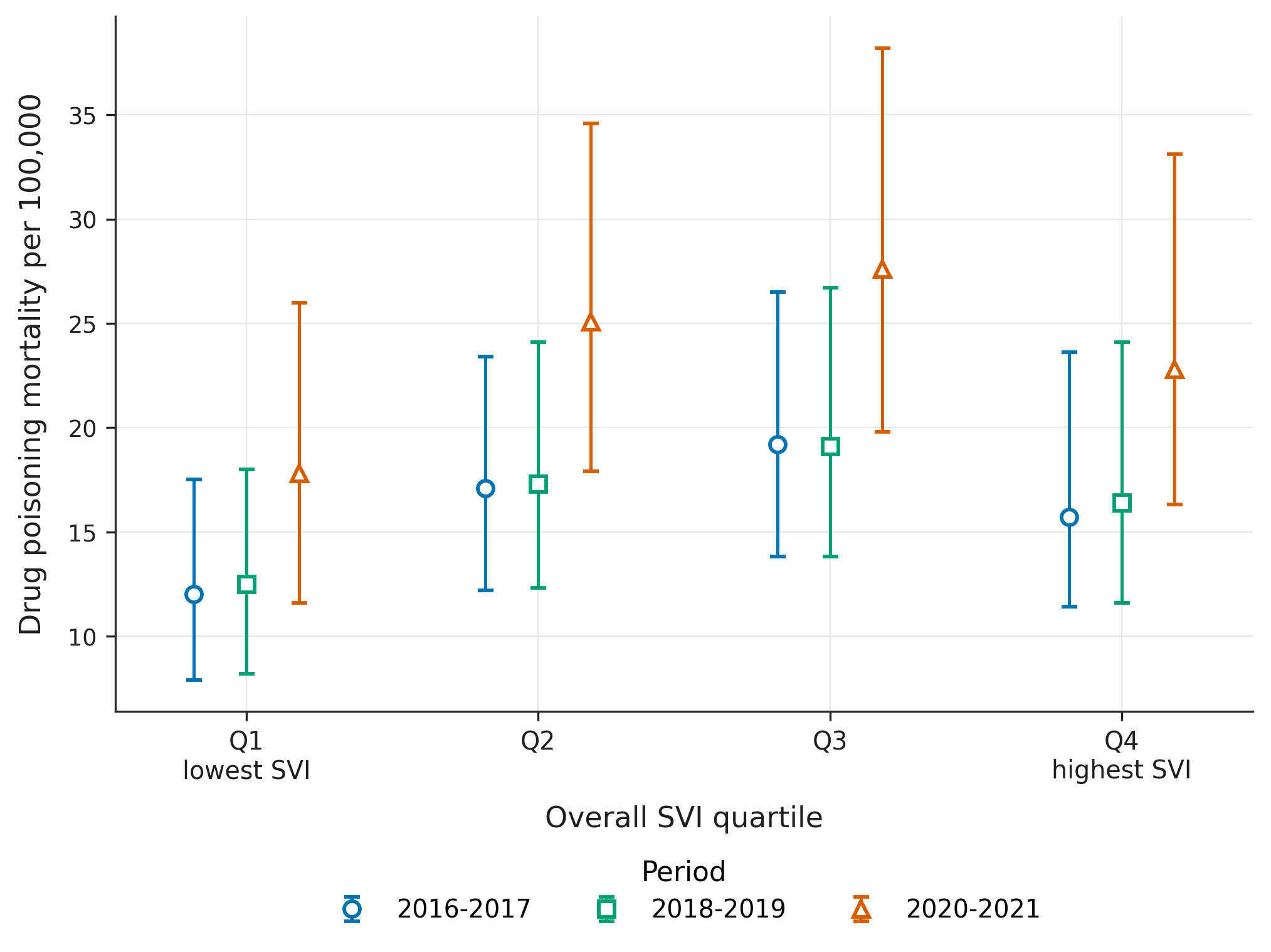}
\caption{Drug poisoning mortality by overall SVI quartile and period.}
\label{fig:des_dpm_svi_quatile_period}
\end{figure}

Figure \ref{fig:des_dpm_svi_quatile_period} shows two-year pooled drug poisoning mortality per 100,000 population across overall SVI quartile in each window. 
Across all three periods, counties in the least vulnerable quartile (Q1) had the lowest median DPM, while middle and upper SVI quartiles experienced higher mortality. However, this gradient was non-monotonic. In 2016–2017, median DPM increased from 12.0 deaths per 100,000 in Q1 to 17.1 in Q2 and 19.2 in Q3, before declining to 15.7 in Q4. A similar pattern was observed in 2018–2019, with median DPM values of 12.5, 17.3, 19.1, and 16.4 across Q1 through Q4, respectively.
The same non-monotonic pattern was observed in the pandemic-era window (2020–2021). Mortality rose sharply across all SVI quartile, with median DPM reaching 17.8 in Q1, 25.1 in Q2, 27.6 in Q3, and 22.8 in Q4.
Within-quartile variability was also greater in 2020-2021, particularly in Q3 (IQR, 19.8–38.2 deaths per 100,000 vs 13.8–26.5 in 2016–2017 and 13.8–26.7 in 2018–2019).
Overall, the non-monotonic pattern, with the highest median DPM in Q3 rather than Q4, was consistent across all three analytic windows.

\begin{figure}[ht]
\centering
\includegraphics[width=0.8\textwidth]{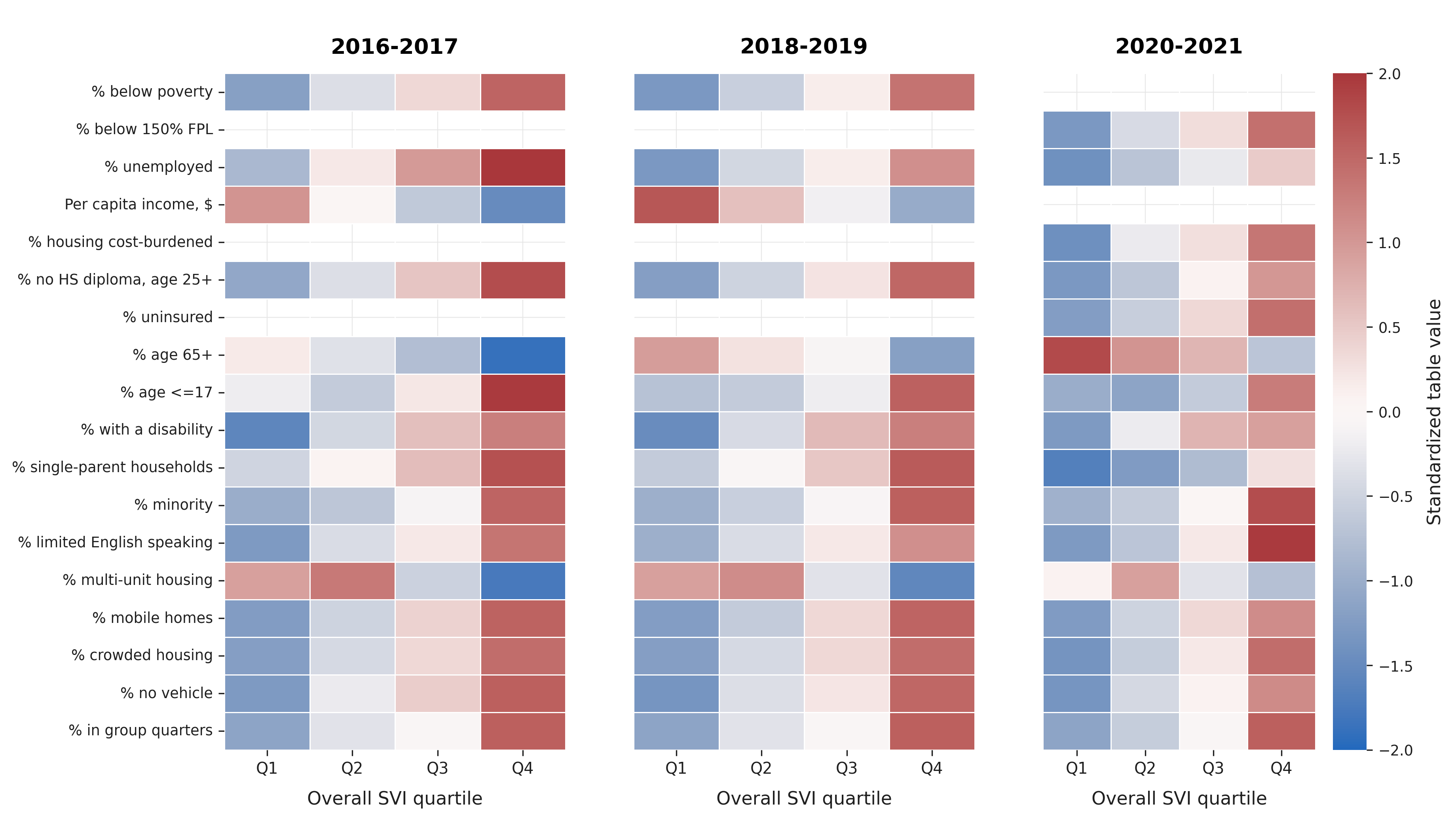}
\caption{Heatmap of SVI subfeature profile by theme, quartile, and period.}
\label{fig:des_svi_theme_subfeature_heatmap}
\end{figure}

Figure \ref{fig:des_svi_theme_subfeature_heatmap} displays a heatmap of standardized subfeature values across SVI quartiles in each of the three periods. As expected from the construction of the overall index, most subfeatures exhibit a monotonic Q1-to-Q4 gradient, and these quartile profiles remain visually stable across all three analytic windows. 
However, two notable departures from monotonicity warrant attention. First, the percentage of residents aged 65 years and over declines from Q1 to Q4, indicating that the least vulnerable counties carry the oldest age distributions on average. Second, the percentage of housing units in multi-unit structures peaks in Q2, showing a non-monotonic pattern across SVI quartiles.

\subsection{Association Analysis}

Among the three candidate models, XGBoost achieved the lowest test-set RMSE in all three analytic windows, with values of 0.304, 0.293, and 0.304, respecttively (Table \ref{tab:model_comparison}). XGBoost also achieved the highest $R^2$ and lowest MAE in each window, yielding the same model ranking across all three performance metrics. XGBoost was therefore selected for SHAP-based interpretation. Linear regression showed lower predictive performance than both tree-based models in each window, consistent with its restrictive linear-additive specification.

\begin{table}[ht]
\centering
\caption{Predictive performance of candidate models in the associational analysis.}
\label{tab:model_comparison}
\begin{tabular}{llccc}
\toprule
Period & Metric & Linear Regression & XGBoost & Random Forest \\
\midrule
\multirow{3}{*}{2016-2017}
  & $R^2$  & 0.490 & \textbf{0.667} & 0.657 \\
  & RMSE   & 0.376 & \textbf{0.304} & 0.309 \\
  & MAE    & 0.299 & \textbf{0.241} & 0.245 \\
\midrule
\multirow{3}{*}{2018-2019}
  & $R^2$  & 0.547 & \textbf{0.700} & 0.680 \\
  & RMSE   & 0.360 & \textbf{0.293} & 0.303 \\
  & MAE    & 0.289 & \textbf{0.230} & 0.241 \\
\midrule
\multirow{3}{*}{2020-2021}
  & $R^2$  & 0.509 & \textbf{0.664} & 0.647 \\
  & RMSE   & 0.368 & \textbf{0.304} & 0.312 \\
  & MAE    & 0.293 & \textbf{0.241} & 0.249 \\
\bottomrule
\end{tabular}
\end{table}

\begin{figure}[ht]
\centering
\includegraphics[width=0.8\textwidth]{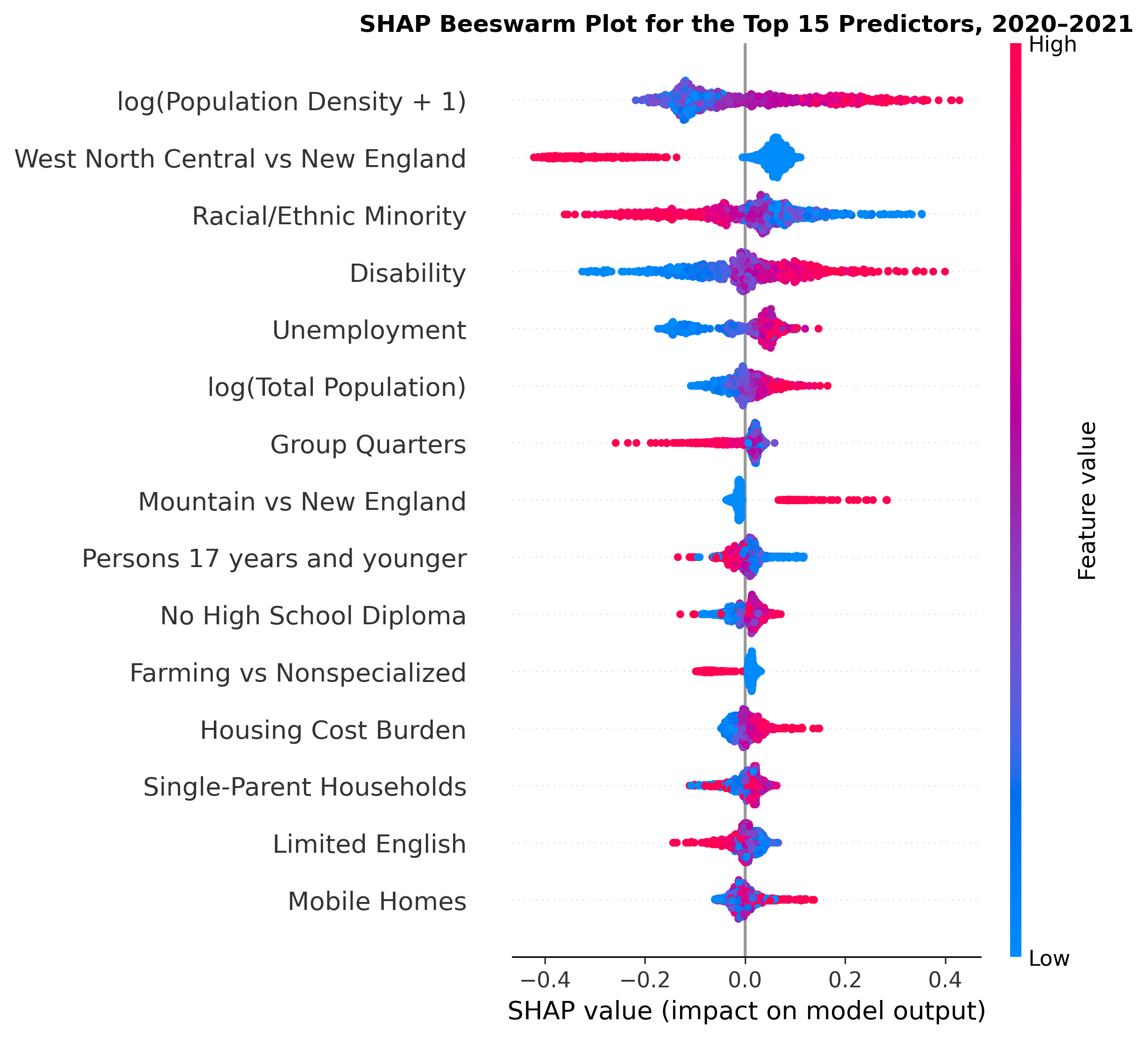}
\caption{SHAP beeswarm plot of top 15 features for XGBoost during 2020-2021.}
\label{fig:beeswarm_top15_2020_2021_XGBoost}
\end{figure}

Figure \ref{fig:beeswarm_top15_2020_2021_XGBoost} presents the SHAP Beeswarm plot of the 15 predictors with the largest mean absolute SHAP values for 2020–2021. Corresponding plots for 2016–2017 and 2018–2019 are shown in Appendix Figures \ref{fig:beeswarm_top15_2016_2017_XGBoost} and \ref{fig:beeswarm_top15_2018_2019_XGBoost}. 
Among the non-SVI covariates, population density and the West North Central census division indicator were the two highest-ranked predictors in all three windows.
Higher population density was consistently associated with higher predicted DPM, with Spearman correlations of 0.87, 0.87, and 0.90 in 2016–2017, 2018–2019, and 2020–2021, respectively. The West North Central division showed strongly negative SHAP contributions relative to the New England reference, while the Mountain division contributed positive ones. These patterns in population density and geographic indicators were consistent across all analytic windows.

Among the SVI sub-features, racial and ethnic minority composition, disability, and unemployment had the largest mean absolute SHAP values in every window. Higher disability ranks were associated with increasingly positive SHAP contributions, with Spearman $\rho$ of 0.97, 0.96, and 0.95 across the three periods. Unemployment showed the same positive direction, with $\rho$ of 0.77, 0.85, and 0.87. Conversely, Racial and ethnic minority composition exhibited the reverse pattern, with $\rho$ of -0.87, -0.86, and -0.90, indicating that higher minority ranks were associated with lower predicted DPM. 
Group quarters and the younger age components also showed negative directional associations with predicted mortality. 
Appendix Tables \ref{tab:shap_direction_svi}, \ref{tab:shap_direction_cov}, and \ref{tab:shap_direction_ind} report the mean absolute SHAP values and directional summaries for all features across the three windows.

\begin{figure}[ht]
\centering

\begin{subfigure}{0.7\textwidth}
    \centering
    \includegraphics[width=\textwidth]{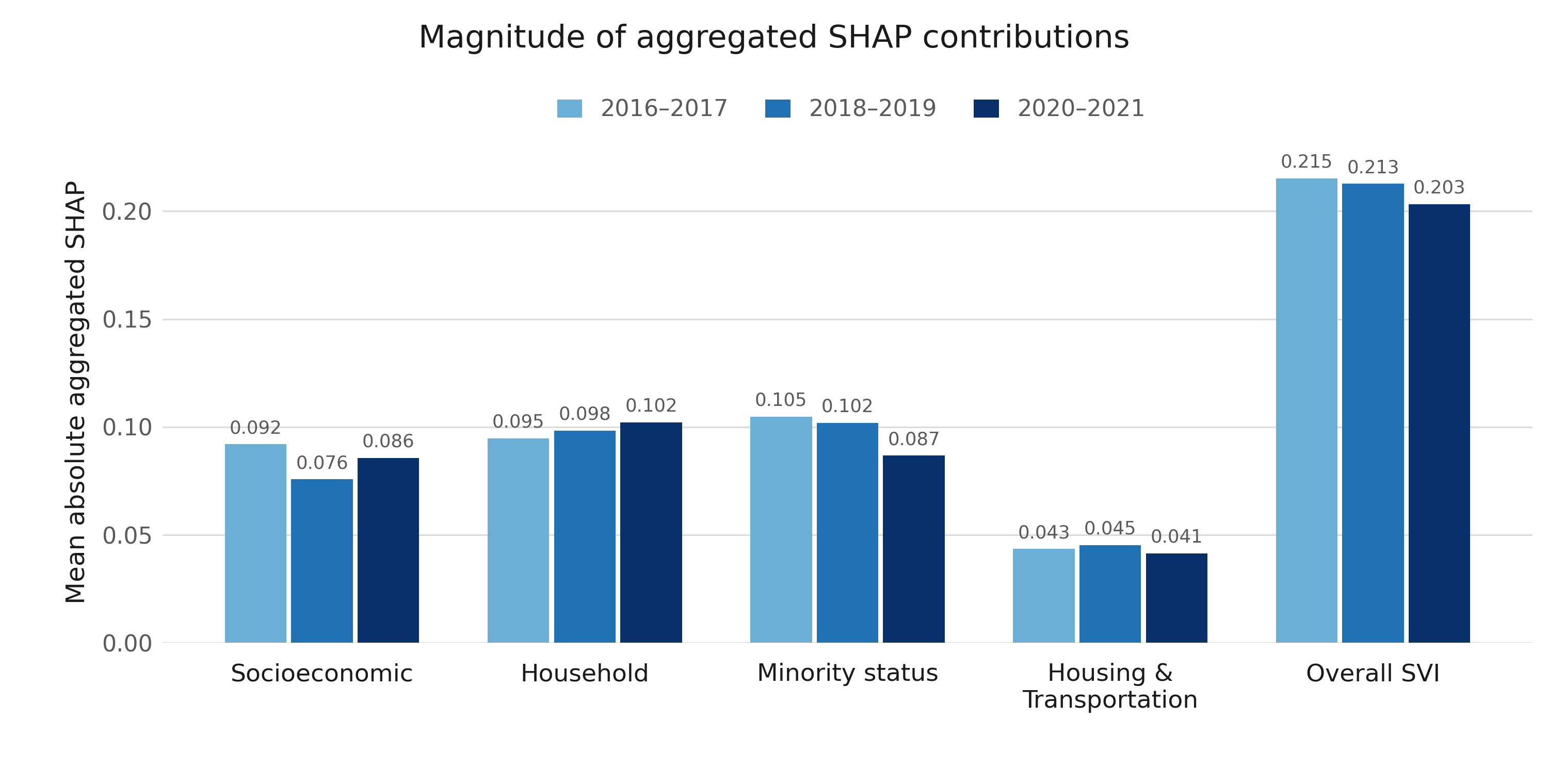}
    \caption{}
    \label{fig:aggregated_magnitude}
\end{subfigure}

\vspace{0.3cm}

\begin{subfigure}{0.7\textwidth}
    \centering
    \includegraphics[width=\textwidth]{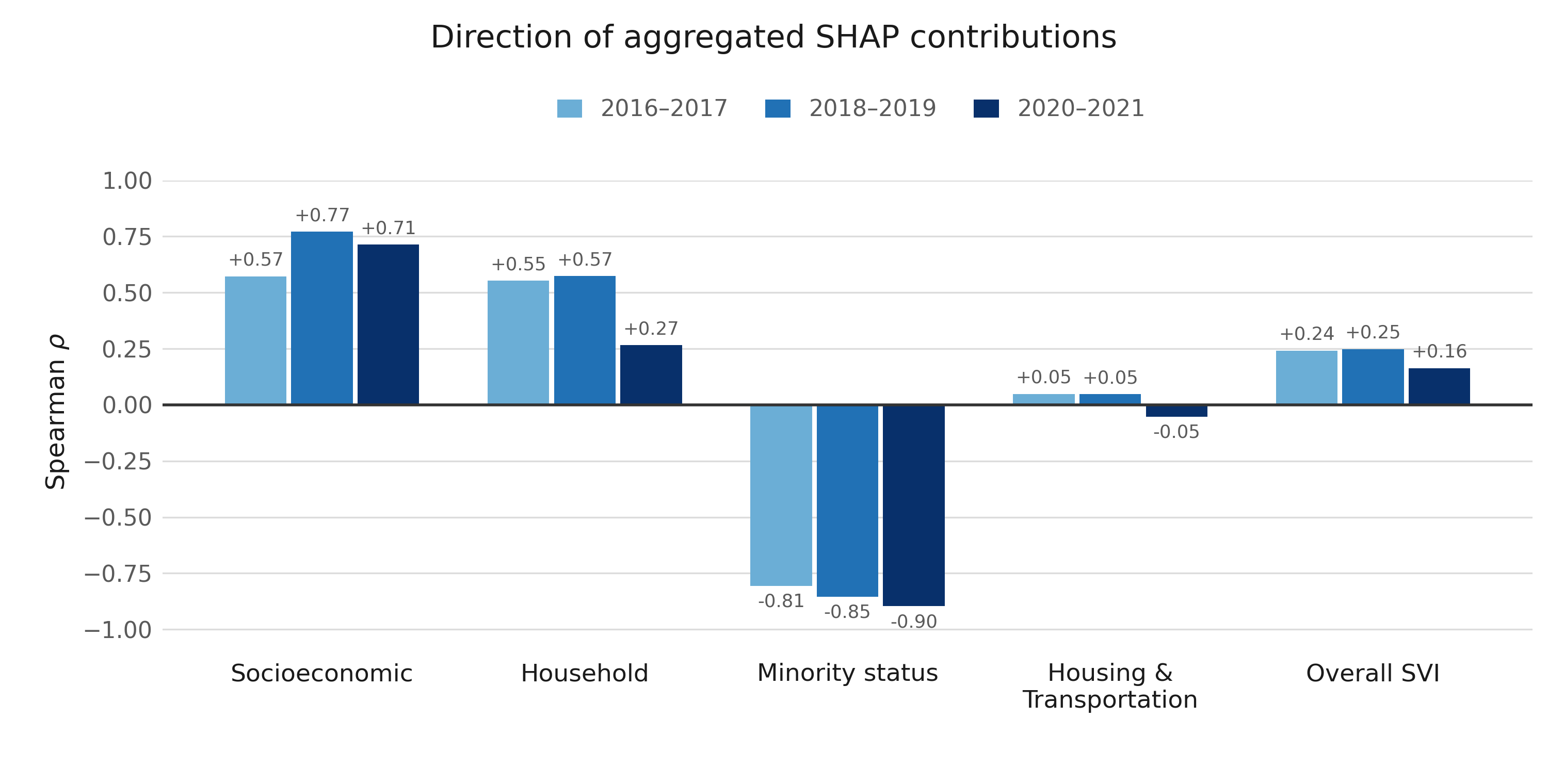}
    \caption{}
    \label{fig:aggregated_direction}
\end{subfigure}

\caption{Magnitude and direction of SVI contributions to predicted drug poisoning mortality. 
Panel \ref{fig:aggregated_magnitude} shows the mean absolute aggregated SHAP contribution for each SVI theme and for overall SVI. 
Signed subfeature-level SHAP values were first summed within each county before absolute values were averaged across counties. 
Panel \ref{fig:aggregated_direction} shows the Spearman correlation between each theme or overall SVI ranking and its corresponding aggregated SHAP contribution, summarising monotonic direction.}

\label{fig:5measures_XGBoost}
\end{figure}

Figure \ref{fig:5measures_XGBoost} illustrates the magnitude and direction of SHAP contributions aggregated from the subfeature model to the SVI theme and overall SVI.
The socioeconomic, household, and minority-status themes had broadly similar mean absolute SHAP magnitudes, while housing and transportation was consistently smaller.
After the signed theme contributions were aggregated within counties, the mean absolute overall-index contribution was 34\%-36\% lower than the sum of the four theme-specific mean absolute contributions. This difference quantifies the extent to which positive and negative theme contributions offset one another.

The directional patterns differed across themes. Higher socioeconomic vulnerability consistently contributed to higher predicted mortality, while the household theme showed a similar positive pattern in the first two windows but weakened in 2020–2021, with Spearman's $\rho$ decreasing from 0.55 and 0.57 to 0.27. Minority Status showed the opposite pattern and remained strongly negative across all three windows, with $\rho = -0.81, -0.85, \text{and} -0.90$. Housing and transportation showed essentially no monotonic directional association, with Spearman’s $\rho$ near zero in all three windows. The corresponding correlations for the aggregated overall-index contribution were positive but modest at 0.24, 0.25, and 0.16. Appendix Table \ref{tab:theme_shap_magnitude} reports the magnitude and direction for each theme and the aggregated overall-index summary.

In the overall-SVI association models, higher overall SVI was consistently associated with higher predicted log DPM and ranked as one of the most influential predictors across all three analytic windows. Corresponding SHAP beeswarm plots are shown in Appendix Figures \ref{fig:overall_svi_shap_2016_2017, fig:overall_svi_shap_2018_2019, fig:overall_svi_shap_2020_2021}

To assess sensitivity to the inclusion of non-SVI covariates, we re-estimated the XGBoost models using only the SVI subfeatures. The principal directional patterns were retained across all three windows. Disability and unemployment remained positively associated with predicted mortality, whereas racial and ethnic minority composition and group quarters remained negatively associated. Full SHAP results are reported in Appendix Figures \ref{fig:beeswarm_svi_2020_2021_XGBoost}, \ref{fig:beeswarm_svi_2018_2019_XGBoost}, and \ref{fig:beeswarm_svi_2016_2017_XGBoost}.

We also repeated the analysis using untransformed DPM as the outcome while retaining the full predictor set. The leading SVI subfeatures and their directional patterns were consistent with the primary log-scale results across all three windows. Corresponding SHAP results are shown in Appendix Figures \ref{fig:rawdpm_beeswarm_svi_2020_2021_XGBoost}, \ref{fig:rawdpm_beeswarm_svi_2018_2019_XGBoost}, and \ref{fig:rawdpm_beeswarm_svi_2016_2017_XGBoost}.

\subsection{Double Machine Learning Analysis}
The causal analysis estimated the adjusted effect of overall SVI on log DPM. We first report the nuisance-model performance and overlap diagnostics, followed by an assessment of the functional form and the estimated exposure-response relationship.

Table \ref{tab:nuisance-performance} summarizes the cross-fitted predictive performance of Elastic net, XGBoost, and Random Forest for the outcome and treatment nuisance functions. XGBoost achieved the lowest RMSE for both nuisance functions in all three analytic windows and was therefore used for both nuisance functions in the DML specification. 

\begin{table}[htbp]
\centering
\caption{Predictive performance of nuisance models across time periods}
\label{tab:nuisance-performance}
\renewcommand{\arraystretch}{1.25}
\begin{tabular}{lllccc}
\hline
Period & Nuisance Role & Metric & Elastic net & XGBoost & Random Forest \\
\hline
\multirow{6}{*}{2016-2017}
& \multirow{3}{*}{Treatment}
& $R^2$ & 0.407 & 0.489 & 0.474 \\
& & RMSE & 0.222 & \textbf{0.206} & 0.209 \\
& & MAE  & 0.181 & 0.164 & 0.167 \\
\cline{2-6}
& \multirow{3}{*}{Outcome}
& $R^2$ & 0.352 & 0.459 & 0.445 \\
& & RMSE & 0.427 & \textbf{0.390} & 0.395 \\
& & MAE  & 0.339 & 0.310 & 0.314 \\
\hline

\multirow{6}{*}{2018-2019}
& \multirow{3}{*}{Treatment}
& $R^2$ & 0.399 & 0.466 & 0.458 \\
& & RMSE & 0.224 & \textbf{0.211} & 0.213 \\
& & MAE  & 0.183 & 0.169 & 0.171 \\
\cline{2-6}
& \multirow{3}{*}{Outcome}
& $R^2$ & 0.349 & 0.448 & 0.447 \\
& & RMSE & 0.427 & \textbf{0.393} & 0.394 \\
& & MAE  & 0.339 & 0.312 & 0.312 \\
\hline

\multirow{6}{*}{2020-2021}
& \multirow{3}{*}{Treatment}
& $R^2$ & 0.404 & 0.470 & 0.469 \\
& & RMSE & 0.223 & \textbf{0.210} & 0.211 \\
& & MAE  & 0.183 & 0.170 & 0.170 \\
\cline{2-6}
& \multirow{3}{*}{Outcome}
& $R^2$ & 0.348 & 0.457 & 0.443 \\
& & RMSE & 0.428 & \textbf{0.390} & 0.396 \\
& & MAE  & 0.339 & 0.309 & 0.314 \\
\hline
\end{tabular}
\end{table}

\begin{figure}[htbp] 
\centering 
\includegraphics[width=\textwidth]{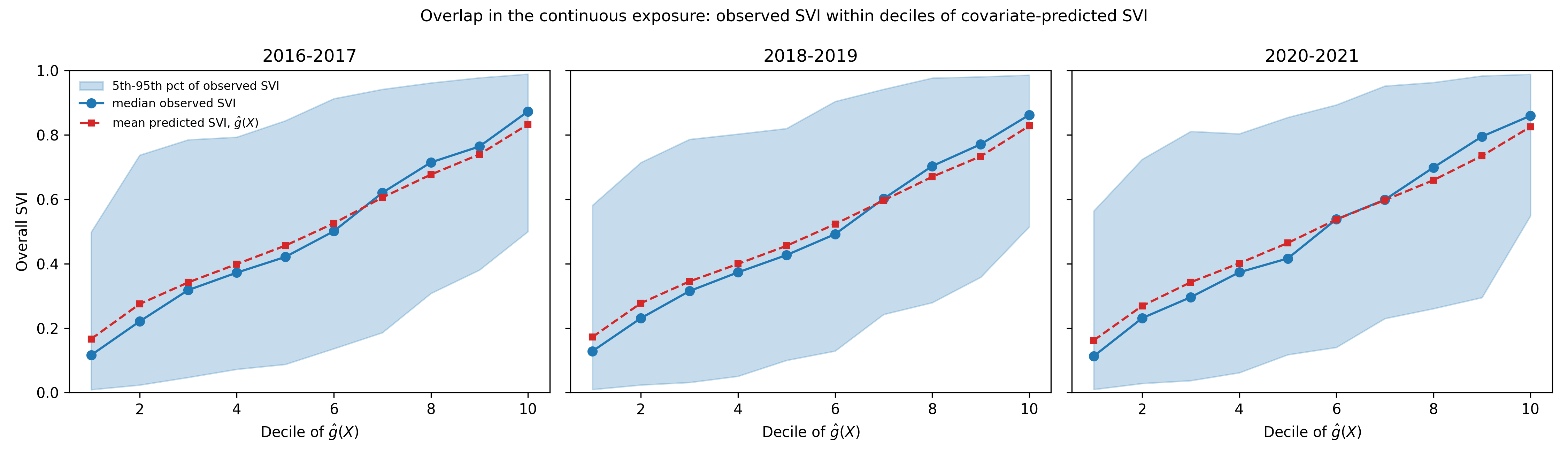} \caption{Overlap in the continuous exposure. Observed SVI distributions are shown within deciles of cross-fitted covariate-predicted SVI, $\hat g(X)$; points indicate medians and shaded bands the 5th–95th percentile range.} 
\label{fig:dml_overlap} 
\end{figure}

After conditioning on $X$, the ratio of residualized to observed SVI standard deviations was 0.72–0.73 across the three analytic windows. Figure \ref{fig:dml_overlap} further shows that observed SVI remained broadly distributed within deciles of covariate-predicted SVI, supporting empirical overlap across the observed exposure range.


The partialling-out plot for the 2020-2021 linear benchmark is shown in Appendix Figure~\ref{fig:dml_residual_scatter_2020_2021}. After residualizing both log DPM and overall SVI with respect to the adjustment set, the residual-on-residual slope was positive, corresponding to the linear DML estimate of $\hat{\theta}=0.259$.

The linearity restriction was rejected in all three analytic windows. For the primary four-knot restricted cubic spline, joint Wald statistics for the nonlinear terms were 130.0, 123.6, and 130.1 in 2016–2017, 2018–2019, and 2020–2021, respectively, with all $p<10^{-26}$.
The fitted exposure-response curves showed a rise-and-decline pattern, with estimated peaks at SVI values of 0.65, 0.65, and 0.63 across the three windows. The nonlinear specification therefore provided the primary description of the causal exposure–response relationship.

\begin{figure}[htbp]
\centering
\includegraphics[width=\textwidth]{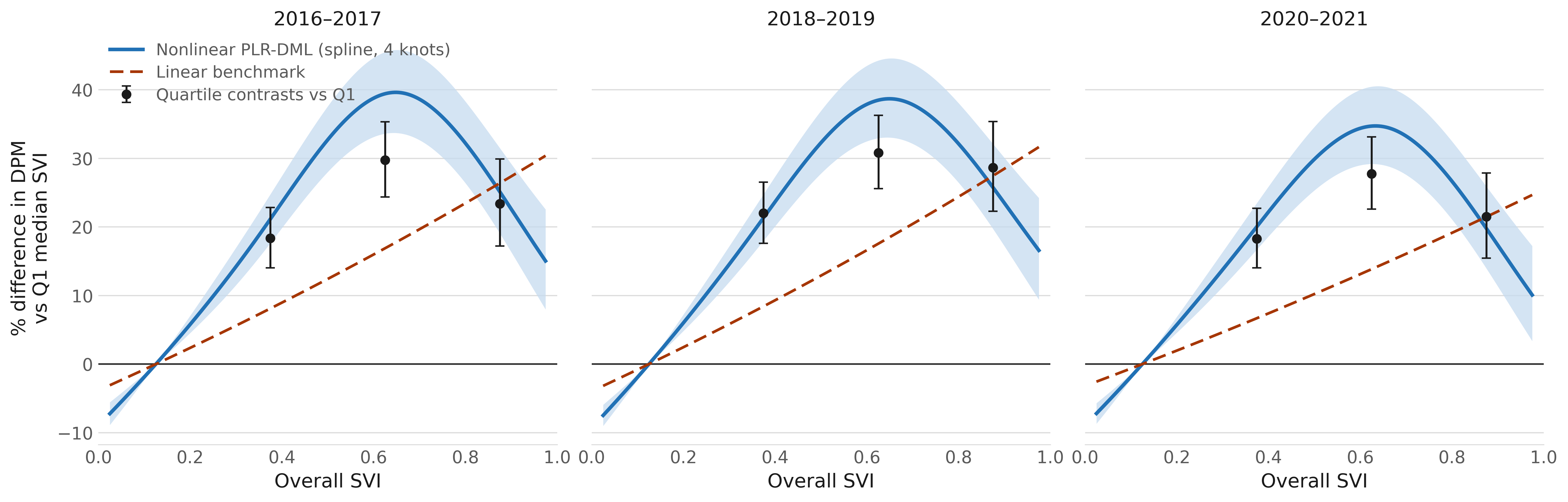}
\caption{Adjusted nonlinear relationship between overall SVI and drug poisoning mortality. The blue curve shows the nonlinear DML estimate (four-knot restricted cubic spline) with shaded 95\% confidence intervals. Black points indicate corresponding quartile contrasts (with 95\% CIs) positioned at each quartile's empirical median. The dashed red line shows the linear DML benchmark.}
\label{fig:dml_nonlinear}
\end{figure}

Figure \ref{fig:dml_nonlinear} shows the fitted nonlinear exposure-response relationship over the central 95\% of the observed SVI distribution, with quartile contrasts overlaid. 
Relative to a county at the median of the first quartile, the continuous spline increased through through the middle of the index, peaked near an SVI of 0.65, and then declined. The quartile models showed the same non-monotonic pattern. DPM in Q3 was estimated to be 29.7\%, 30.8\%, and 27.8\% higher than in Q1 in 2016–2017, 2018–2019, and 2020–2021, respectively, with all $p < 0.001$. 
Following this peak, mortality declined in the fourth quartile. This downward turn was statistically significant in 2016–2017 ($-4.9\%$, $p = 0.029$) and 2020–2021 ($-4.9\%$, $p = 0.023$), but not in 2018–2019 ($-1.7\%$, $p = 0.46$).

The upper-tail decline was more clearly resolved in the quintile models, which provided finer separation across the higher end of the SVI distribution.
Adjusted mortality peaked in the fourth quintile and was significantly lower in the fifth across all three windows, by 10.4\%, 9.1\%, and 8.1\%, respectively. Appendix Figure \ref{fig:dml_nonlinear_quintile} shows the fitted curve with quintile contrasts overlaid. In 2018–2019, the Quintile 5 versus Quintile 4 contrast was significant even though the corresponding Quartile 4 versus Quartile 3 contrast in the quartile model was not. The quintile results were consistent with the upper-tail decline observed in the continuous spline.

For comparison, Table \ref{tab:dml_results} reports the linear PLR-DML benchmark estimates. The estimates were positive and statistically significant ($p < 0.001$) across all three windows, corresponding to average adjusted mortality increases of approximately 2.6\% to 3.3\% per 0.10-unit increment of overall SVI. These estimates summarize an average linear trend across the SVI distribution and do not capture the nonlinear shape established above.
Across the interquartile range from SVI 0.25 to 0.75,  the linear benchmark implied mortality differences of approximately 14\% to 18\%, compared with 19\% to 24\% from the nonlinear model.
Appendix Table \ref{tab:decile_nonlinear_vs_linear} compares the nonlinear and linear estimates at common SVI values across the exposure range. When evaluated at the same SVI values relative to the Q1-median reference, the linear benchmark yielded smaller mortality contrasts than the nonlinear model through the middle of the distribution and larger contrasts in the upper tail. The two fitted relationships crossed between approximately 0.8 and 0.9.

\begin{table}[htbp]
\centering
\caption{Linear PLR-DML benchmark estimates for overall SVI and drug poisoning mortality.}
\label{tab:dml_results}
\renewcommand{\arraystretch}{1.25}
\begin{tabular}{lllllllll}
\hline
Period & Outcome & Treatment & $\hat{\theta}$ & SE & 95\% CI & p-value &
\begin{tabular}[c]{@{}l@{}}\% change\\ per 0.1 SVI\end{tabular} &
\begin{tabular}[c]{@{}l@{}}95\% CI,\\ \%\end{tabular} \\
\hline
2016-2017 & log DPM & SVI & 0.312 & 0.035 & [0.245, 0.380] & $<0.001$ & 3.17\% & [2.48\%, 3.87\%] \\
2018-2019 & log DPM & SVI & 0.324 & 0.034 & [0.256, 0.391] & $<0.001$ & 3.29\% & [2.60\%, 3.98\%] \\
2020-2021 & log DPM & SVI & 0.259 & 0.035 & [0.192, 0.327] & $<0.001$ & 2.63\% & [1.94\%, 3.33\%] \\
\hline
\end{tabular}
\end{table}


Sensitivity analyses of the nonlinear specification are reported in Appendix Table \ref{tab:sensitivity_nonlinear}. The rise-and-decline pattern was preserved across alternative spline specifications, nuisance learners, and geographic adjustment. Across the three-, four-, and five-knot specifications, the estimated peaks remained between SVI values of 0.58 and 0.69.
Linearity was rejected in every specification, and the fifth quintile remained significantly below the fourth in all three windows. The spline contrast between SVI values of 0.25 and 0.75 also remained positive across specifications.

The upper-tail decline was attenuated, but not eliminated, when census division was replaced by state indicators. The magnitude of the negative Q5–versus-Q4 contrast was reduced from between $-8.1\%$ and $-10.4\%$ in the primary models to between $-4.8\%$ and $-6.1\%$ under state-level adjustment, but remained statistically significant across all three windows. 
Residualized SVI retained 65\%–68\% of its original standard deviation, and 57\%–60\% of the total SVI variation occurred within states. These results suggest that between-state geographic composition contributes to the magnitude of the upper-tail decline but does not fully account for it.

The nonlinear finding was also robust to the outcome transformation. Re-estimating the models using untransformed DPM preserved the rise-and-decline pattern. Linearity remained rejected, the fitted maxima occurred at similar SVI values, and mortality in the fifth quintile remained significantly below the fourth across all three windows (Appendix Table \ref{tab:sensitivity_rawdpm_nonlinear}). 
The corresponding check on the linear benchmark is reported separately in Appendix Table \ref{tab:sensitivity_rawdpm_linear}, and linear benchmark under each alternative nuisance learner appears in the final column of Appendix Table \ref{tab:sensitivity_nonlinear}.



%% file: Discussion.tex
\section{Discussion}

Overall social vulnerability was positively associated with county-level drug poisoning mortality (DPM) across its distribution, but the adjusted exposure-response relationship declined in the upper tail. 
At the SVI theme levels, socioeconomic vulnerability and disability contributed positively to predicted mortality, minority-status composition contributed negatively, and housing and transportation contributed little. 
These findings were consistent in all three windows. For overall SVI, the upper-tail decline and differing component directions recurred consistently in all three windows. 
These findings emerged from two complementary analyses applied to the same counties and time periods.

The non-monotonic pattern was already visible before adjustment. Median mortality rose from the least vulnerable counties through the third quartile in all three windows and then fell in the fourth quartile. 
Population density may partly contribute to this pattern, because it peaked in the middle SVI quartiles and was among the strongest positive predictors in the associational analysis.
Geographic composition may also contribute to this pattern. Zhang and Monnat found that associations between socioeconomic and household vulnerability and drug-poisoning mortality varied substantially across regions, with positive associations concentrated in Appalachia and selected Midwestern and Plains areas, while some high-distress areas in the Mississippi Delta and Gulf Coast showed no significant association with drug-poisoning mortality \cite{zhang2024geographically}.
The same rise-and-decline shape persisted after adjustment for density, urbanicity, and broad geographic location. 
Minority composition rises steeply across the top of the index and contributes negatively. Because the minority-status indicators are constitutive components of the exposure, they are not included in the adjustment set; conditioning on the adjustment set  would fundamentally alter the estimand. Consequently, the upper-tail decline may partly reflect this shifting component composition among the most vulnerable counties, rather than a diminishing hazard of vulnerability itself. The SHAP analysis and the DML exposure-response model are separate specifications rather than a unified decomposition, so this remains an interpretive synthesis rather than a formally established mechanism. Geographic composition contributes as well: Replacing census divisions with state indicators roughly halves the upper-tail decline, while it remained statistically significant in all three windows.

The socioeconomic theme gave the steadiest positive signal across all three windows, consistent with a large number of study linking county poverty, unemployment, and economic distress to higher overdose death rates \cite{ghertner2018opioid, monnat2018factors, fink2024social}. The household theme remained important in magnitude, although its positive directional association weakened substantially in 2020-2021. Housing and transportation contributed relatively little, with near-zero directional correlations across all three periods. Higher racial and ethnic minority composition was consistently associated with lower predicted mortality. The differing theme-level directions show that the overall SVI can obscure component-level heterogeneity. Stokes et al. similarly found divergent associations among individual SVI variables that could appear weak or null after aggregation at the theme level \cite{stokes2023county}.

County-level negative correlation between minority composition and predicted DPM should not be interpreted as a protective effect of minority status. After accounting for other predictors, minority composition more plausibly functions as a marker of differential geographic exposure. 
Geographic exposure changed substantially during the fentanyl era. Powell et al. estimated that differential geographic exposure accounted for roughly 40\% of the additional growth in Black relative to White opioid-overdose mortality from 2010 to 2020 \cite{powell2025geography}. 
Furthermore, structural racism operates partly through channels independent of socioeconomic status, including residential segregation, hyper-incarceration, and everyday discrimination \cite{zhu2024structural}. The SVI minority-status theme measures none of these mechanisms, although they are measurable and consequential at finer spatial scales \cite{uzzi2025fatal}.

Within the themes, higher disability and higher unemployment ranks were the strongest positive contributors.
Both age components and the group quarters component showed negative SHAP directions. Similar inverse patterns have been reported for nonfatal overdose outcomes, with higher county shares of older or younger residents and of residents living in group quarters associated with lower overdose rates \cite{stokes2023county}. For fatdrug poisoning mortality, group-quarters prevalence has also been reported to be inversely associated with mortality \cite{frankenfeld2019county, fink2024social}.
Disability is the most consistent single result, holding its direction in every window and surviving the SVI-only specification.
This positive disability pattern is consistent with elevated opioid-overdose mortality among Medicare beneficiaries with disabilities \cite{kuo2019association} and with county-level spatial overlap between disability and opioid-related mortality \cite{deas2025identifying}. Opioid use is also common and often chronic among disabled Medicare beneficiaries \cite{morden2014prescription}.
The positive unemployment contribution is consistent with hypotheses linking economic distress to overdose mortality through chronic stress, despair, and self-medication \cite{monnat2019contributions} and the inverse age directions fit the lower baseline risk of the elderly and the very young \cite{el2023association}. 
Summing the four theme contributions reduced the mean absolute attribution for the overall index by roughly a third, so the composite's directional signal is markedly weaker than that of its strongest components. Grobman et al. report a comparable attenuation over 2011-2020 \cite{grobman2026social}, and Stokes et al. caution that reading the SVI only at the overall or theme level can hide divergent component associations \cite{stokes2023county}.

The DML analysis quantified the adjusted exposure-response relationship for overall SVI. Under the identifying assumptions of conditional exchangeability, positivity, and consistency, the estimated partialling-out contrasts can be interpreted as causal effects of differences in overall SVI \cite{chernozhukov2018double,moccia2024machine}. 
The linear benchmark estimated a positive average causal effect of overall SVI on log DPM across all three periods. The nonlinear specification refined this result by showing that the effect was not constant across the SVI distribution: counties at the 75th percentile of overall SVI were estimated to have 19\% to 24\% higher DPM than they would at the 25th percentile, while the exposure-response curve flattened and eventually declined in the upper tail. Thus, the positive coefficient from the linear model summarizes the relationship across the full SVI range, whereas the nonlinear model reveals substantial variation in the causal contrast across exposure levels. The rise-and-decline pattern was reproduced across alternative spline bases, geographic controls, nuisance learners, and outcome scales, with the fifth quintile below the fourth throughout and the upper-tail decline ranging from roughly 5\% to 15\% across specifications.
Residualized-exposure diagnostics further showed that substantial variation in overall SVI remained after conditioning on the adjustment set, with approximately three-quarters of the original variation retained. 
Taken together, the linear and nonlinear DML analyses indicate a positive average causal relationship between overall SVI and DPM while showing that this relationship is heterogeneous across the SVI distribution and reverses direction at the highest exposure levels.

The three windows bracket a major shock to the overdose epidemic, yet the exposure-response relationship remained broadly similar across them. Median mortality rose by roughly 45\% in every vulnerability quartile between 2016-2017 and 2020-2021, and within-quartile variation widened. Despite this level shift, linearity was rejected in all three windows and the fitted maximum remained near an SVI of 0.65. The estimated interquartile contrast was somewhat smaller in 2020-2021 (18.9\%) than in the two pre-pandemic windows (23.8\% and 23.0\%), although the confidence intervals overlapped substantially. What changed more clearly was the directional contribution of individual vulnerability dimensions, particularly the household theme, whose predictive magnitude remained substantial while its positive directional association weakened. This pattern may reflect greater heterogeneity in the pathways linking household characteristics to overdose during the early pandemic. Model-based apportionment of the 2020 increase attributed 43\% of estimated excess opioid overdose deaths to growth in fentanyl exposure and a further 8\% to declines in methadone and injectable naltrexone treatment \cite{stringfellow2023enumerating}. 
Emergency medical services data from early 2020 also documented disruptions in care engagement, including increased refusal of transport after the declaration of a state of emergency \cite{slavova2020signal,walters2022structural}. 
These pandemic-specific processes need not follow the ordering captured by the SVI household theme. The persistence of the overall SVI pattern may partly reflect the structural socioeconomic and demographic conditions captured by the index, many of which change more gradually than acute features of the overdose environment \cite{cdc2020svi}. Grobman et al. reported weakening of the overall SVI-mortality association as mortality increased over 2011-2020 \cite{grobman2026social}; the three two-year windows examined here are better suited to assessing short-term stability than attenuation over that longer horizon.

The results bear on how vulnerability measures are interpreted when allocating prevention resources. They suggest that component-specific measures, particularly the socioeconomic dimension and the disability sub-feature, may align more closely with observed mortality burden than the overall composite. While the overall SVI index points in roughly the right direction over most of its range, it does not order counties correctly at the top of its own scale. Consequently, jurisdictions may benefit from identifying counties with high scores on these specific components rather than relying exclusively on the overall SVI composite to target interventions. Beyond measurement, these findings have broader policy implications. The prominence of the socioeconomic pathway, which runs partly through housing instability where rising homelessness tracks increases in accidental poisoning mortality \cite{bradford2024higher}, suggests that upstream economic supports can serve as vital complements to established downstream strategies. Stabilizing economic conditions through housing assistance, employment programs, and income support may effectively address the structural determinants of overdose risk. In practice, these upstream investments should be paired with expanded naloxone distribution, medications for opioid use disorder, harm-reduction services, and disability-informed pain management. Finally, the inverse minority-status direction carries a critical equity implication. 
Because the epidemic’s burden has shifted rapidly toward Black and other minoritized populations, whose overdose mortality surpassed that of White Americans by 2019 \cite{powell2025geography, smith2025widening, cano2025racial}, allocation frameworks built primarily on cross-sectional vulnerability patterns may lag behind the communities where the burden is now growing fastest. 

Pairing area-based vulnerability metrics with timely, race-stratified public health surveillance is essential to close this gap and ensure an equitable, race-conscious approach to overdose prevention and resource distribution.

Several limitations bear on specific conclusions. The analysis is ecological, so the component directions and the adjusted contrast describe geographic patterns and do not transfer to individuals, whose risk concentrates at intersections of race and ethnicity, education, age, and nativity that a county composite cannot resolve \cite{cano2025extreme}. Unmeasured confounders could bias the DML estimates in an indeterminate direction, particularly those capturing local illicit fentanyl penetration and potency, harm reduction and treatment capacity, structural stigma, and individual-level substance use histories. Two sources of uncertainty are also left out of the standard errors. They treat counties as independent observations, so residual spatial dependence remaining after geographic adjustment would leave them understated and the confidence intervals too narrow; and the outcome is a set of model-based NCHS estimates produced by hierarchical Bayesian smoothing, treated here as observed, so their estimation uncertainty is not propagated, while the smoothing itself may attenuate extremes and compress the estimated gradient. The composite nature of the exposure further limits the interpretation in two distinct ways. First, because overall SVI is produced by varying component configurations, the resulting estimate does not capture the effect of altering any individual subfeature. Second, because the ranking is defined relative to all other counties within a given vintage, the estimand reflects relative positional contrasts across the distribution rather than the effect of an intervention that directly modifies a county’s vulnerability score. County aggregation can mask heterogeneity across diverse communities, and the standard urban-rural designations used here impose categorical boundaries that may not track the social processes generating risk \cite{textor2023rethinking}. The 2020 SVI revision restricts cross-window comparison at component level, which is why the temporal comparisons rest on the overall index and the themes, and the window ends in 2021, leaving untested whether the relationship recurs in the period of sustained national decline. Addressing these would require county measures of the local illicit drug supply and of treatment and harm-reduction capacity, estimation stratified by urbanicity and region to test the common-effect assumption that prior stratified work calls into question \cite{monnat2018factors}, and vulnerability dimensions harmonized across SVI releases.

Several extensions could strengthen the analysis. County-level indicators of the local illicit drug supply derived from forensic seizure records, measures of treatment and harm-reduction capacity, and proxies for structural stigma could improve adjustment for important sources of confounding. The analysis also assumes a common effect across counties. Because the descriptive patterns here and prior stratified work \cite{monnat2018factors} both suggest the relationship differs by urbanicity and region, estimating it separately for urban and rural counties or across census divisions would test that assumption and show whether the upper-tail reversal is concentrated in particular settings or general to them. The temporal comparison is descriptive for a related reason: fitting the three windows jointly with an exposure-by-period interaction would convert the overlapping confidence intervals reported here into a test of whether the gradient moved, and bootstrapping the SHAP directional summaries would attach uncertainty to the component shifts, which the present rank correlations lack. Finer spatial resolution, at census-tract or ZIP-code level where data permit, would help separate neighborhood processes from county aggregates. Vulnerability dimensions harmonized across releases, built from consistently defined underlying census variables, would make component-level trends comparable over time, which the 2020 revision currently prevents. Extending past 2021 would establish whether the shape documented here held through the sustained national decline, and whether the dimensions that matter most shifted as the epidemic's geography and drug supply continued to change.

%% file: Conclusion.tex
\section{Conclusion}
We examined county social vulnerability and drug poisoning mortality across 2016-2017, 2018-2019, and 2020-2021, interpreting a predictive model with SHAP and estimating the covariate-adjusted exposure-response relationship with double machine learning.

In the associayional analysis, the socioeconomic and household dimensions tracked higher predicted mortality, with disability and unemployment the strongest positive subfeature contributors, while racial and ethnic minority composition was inversely associated in all three periods. This inverse county-level association does not imply lower individual risk among minority residents.

The adjusted relationship between overall SVI and DPM was nonlinear. Mortality increased through the middle of the SVI distribution and declined in the upper tail, a pattern that persisted across alternative model specifications and outcome scales. These results indicate that overall SVI does not provide a uniformly increasing ordering of overdose mortality across its full range.

The upper-tail decline may reflect differences in county composition and geographic exposure to the overdose epidemic, but its mechanism was not identified. Overall SVI is therefore most informative when interpreted alongside its component dimensions and current overdose surveillance rather than as a standalone risk scale.
\clearpage

%% file: Appendix.tex
\section{Variable Definitions}

\setcounter{table}{0}
\setcounter{figure}{0}
\renewcommand{\thetable}{\thesection.\arabic{table}}
\renewcommand{\thefigure}{\thesection.\arabic{figure}}

\begin{longtable}{@{}p{0.16\textwidth} p{0.22\textwidth} p{0.09\textwidth} p{0.45\textwidth}@{}}
\caption{Variables used in the analysis and their definitions.}\label{tab:variable-definitions} \\
\toprule
Role & Variable & Type & Definition \\
\midrule
\endfirsthead
\multicolumn{4}{c}{{\tablename\ \thetable{} - continued from previous page}} \\
\toprule
Role & Variable & Type & Definition \\
\midrule
\endhead
\midrule
\multicolumn{4}{r}{{Continued on next page}} \\
\endfoot
\bottomrule
\endlastfoot
Outcome & Drug poisoning mortality rate & Continuous & Model-based crude death rates for drug overdose per 100,000 population. \\
 & Log drug poisoning mortality & Continuous & Natural logarithm of the drug poisoning mortality rate. Used as the outcome in models. \\
\addlinespace
Main exposure & Overall Social Vulnerability Index & Continuous & Overall SVI percentile ranking ranging from 0 to 1. Higher values indicate greater social vulnerability. \\
\addlinespace
SVI theme measure & Socioeconomic status & Continuous & SVI theme-level percentile ranking for socioeconomic vulnerability. \\
 & Household composition and disability & Continuous & SVI theme-level percentile ranking for household composition and disability-related vulnerability. \\
 & Minority status and language & Continuous & SVI theme-level percentile ranking for racial or ethnic minority status and language-related vulnerability. \\
 & Housing and transportation & Continuous & SVI theme-level percentile ranking for housing type and transportation-related vulnerability. \\
\addlinespace
SVI component measure & Poverty & Continuous & Percentile ranking for percentage of persons below poverty. Available in 2016 and 2018 SVI releases. \\
 & Poverty below 150\% threshold & Continuous & Percentile ranking for percentage of persons below 150\% of the poverty threshold. Available in the 2020 SVI release. \\
 & Unemployment & Continuous & Percentile ranking for percentage of civilian population age 16 or older who are unemployed. \\
 & Per capita income & Continuous & Percentile ranking for per capita income estimate. Available in 2016 and 2018 SVI releases. \\
 & Housing cost burden & Continuous & Percentile ranking for housing cost burden. Available in the 2020 SVI release. \\
 & No high school diploma & Continuous & Percentile ranking for percentage of persons aged 25 years or older without a high school diploma. \\
 & Uninsured & Continuous & Percentile ranking for percentage of civilian noninstitutionalized population without health insurance. \\
 & Aged 65 or older & Continuous & Percentile ranking for percentage of persons aged 65 years or older. \\
 & Aged 17 or younger & Continuous & Percentile ranking for percentage of persons aged 17 years or younger. \\
 & Disability & Continuous & Percentile ranking for percentage of civilian noninstitutionalized population with a disability. \\
 & Single-parent household & Continuous & Percentile ranking for percentage of single-parent households with children under 18. \\
 & Racial or ethnic minority status & Continuous & Percentile ranking for percentage of persons who are racial or ethnic minorities, defined as all persons except non-Hispanic White. \\
 & Limited English & Continuous & Percentile ranking for percentage of persons age 5 or older who speak English less than well. \\
 & Multi-unit housing & Continuous & Percentile ranking for percentage of housing in structures with 10 or more units. \\
 & Mobile homes & Continuous & Percentile ranking for percentage of housing units that are mobile homes. \\
 & Crowding & Continuous & Percentile ranking for percentage of occupied housing units with more people than rooms. \\
 & No vehicle & Continuous & Percentile ranking for percentage of households without a vehicle. \\
 & Group quarters & Continuous & Percentile ranking for percentage of population living in group quarters. \\
\addlinespace
Geographic covariate & Urban-rural classification & Categorical & NCHS county urban-rural category. \\
 & Census division & Categorical & U.S. Census division. \\
\addlinespace
Demographic covariate & Population density & Continuous & Population per square mile. \\
 & Resident population & Continuous & Resident population. \\
 & Daytime population ratio & Continuous & Ratio of daytime population to resident population. \\
 & Sex ratio & Continuous & Ratio of male to female population. \\
\addlinespace
Economic or structural covariate & County economic type & Categorical & ERS county economic type classification including nonspecialized, farming, mining, manufacturing, federal or state government, and recreation counties. \\
 & Population-loss county & Binary & Indicator for counties classified as population-loss counties. \\
 & Retirement-destination county & Binary & Indicator for counties classified as retirement-destination counties. \\
\end{longtable}

\clearpage

\newgeometry{top=1.2cm, bottom=1.2cm, left=1.5cm, right=1.5cm}
\begin{landscape}
\thispagestyle{empty} 
\centering 
\section{Descriptive Table}

\captionof{table}{County characteristics by Social Vulnerability Index (SVI) quartile across three vintages.}
\label{tab:table1_svi_combined}
\vspace{6pt}                           

\scriptsize
\setlength{\tabcolsep}{3pt}
\renewcommand{\arraystretch}{1.05}
\resizebox{\dimexpr\textheight+\headheight+\headsep+\footskip\relax}{!}{%
\begin{tabular}{@{}p{3cm} ccccc ccccc ccccc@{}}
\toprule
 & \multicolumn{5}{c}{2016-2017} & \multicolumn{5}{c}{2018-2019} & \multicolumn{5}{c}{2020-2021} \\
\cmidrule(lr){2-6} \cmidrule(lr){7-11} \cmidrule(lr){12-16}
Variable & Q1 & Q2 & Q3 & Q4 & $p$ & Q1 & Q2 & Q3 & Q4 & $p$ & Q1 & Q2 & Q3 & Q4 & $p$ \\
\midrule
N counties & 783 & 782 & 782 & 783 & - & 783 & 782 & 782 & 782 & - & 783 & 783 & 781 & 783 & - \\
\addlinespace[2pt]
\multicolumn{16}{@{}l}{\textbf{A. Drug Poisoning Mortality}} \\
\addlinespace[1pt]
\quad DPM (deaths per 100,000) & 12.0 [7.9-17.5] & 17.1 [12.2-23.4] & 19.2 [13.8-26.5] & 15.7 [11.4-23.6] & $<$0.001 & 12.5 [8.2-18.0] & 17.3 [12.3-24.1] & 19.1 [13.8-26.7] & 16.4 [11.6-24.1] & $<$0.001 & 17.8 [11.6-26.0] & 25.1 [17.9-34.6] & 27.6 [19.8-38.2] & 22.8 [16.3-33.1] & $<$0.001 \\
\addlinespace[2pt]
\multicolumn{16}{@{}l}{\textbf{B. SVI / Social Vulnerability}} \\
\addlinespace[1pt]
\multicolumn{16}{@{}l}{\textit{Socioeconomic}} \\
\quad \% below poverty & 10.1 $\pm$ 3.1 & 14.2 $\pm$ 3.6 & 17.7 $\pm$ 3.6 & 23.6 $\pm$ 6.0 & $<$0.001 & 9.6 $\pm$ 3.0 & 13.3 $\pm$ 3.4 & 16.7 $\pm$ 3.8 & 22.8 $\pm$ 6.1 & $<$0.001 & - & - & - & - & - \\
\quad \% below 150\% FPL & - & - & - & - & - & - & - & - & - & - & 16.7 $\pm$ 4.7 & 22.0 $\pm$ 5.6 & 26.3 $\pm$ 5.9 & 32.9 $\pm$ 7.5 & $<$0.001 \\
\quad \% unemployed & 4.3 [3.0-5.5] & 6.2 [5.0-7.5] & 7.6 [6.4-8.9] & 9.6 [7.8-11.6] & $<$0.001 & 3.5 [2.6-4.5] & 5.0 [3.9-6.0] & 6.1 [5.1-7.2] & 7.8 [6.3-9.4] & $<$0.001 & 3.3 [2.3-4.2] & 4.6 [3.8-5.5] & 5.4 [4.4-6.5] & 6.7 [5.4-8.5] & $<$0.001 \\
\quad Per capita income, \$ & 29,086 [26,786-32,222] & 25,393 [23,274-28,098] & 22,966 [20,804-25,406] & 19,842 [17,932-21,749] & $<$0.001 & 31,406 [28,560-34,623] & 27,394 [25,230-30,564] & 24,659 [22,575-27,424] & 21,504 [19,172-23,590] & $<$0.001 & - & - & - & - & - \\
\quad \% housing cost-burdened & - & - & - & - & - & - & - & - & - & - & 19.0 $\pm$ 4.4 & 21.8 $\pm$ 4.6 & 23.0 $\pm$ 4.6 & 25.5 $\pm$ 5.4 & $<$0.001 \\
\quad \% no HS diploma, age 25+ & 8.0 [6.4-9.7] & 11.1 [9.3-13.7] & 15.1 [12.1-18.9] & 20.5 [16.8-23.6] & $<$0.001 & 7.4 [6.0-9.0] & 10.5 [8.8-13.2] & 13.8 [11.4-17.6] & 19.3 [15.7-22.9] & $<$0.001 & 7.0 [5.6-8.8] & 9.8 [7.8-12.4] & 13.1 [10.2-16.5] & 17.1 [13.8-20.8] & $<$0.001 \\
\quad \% uninsured & 7.7 [6.0-10.0] & 10.3 [8.2-13.0] & 12.6 [10.4-15.7] & 15.5 [13.0-19.0] & $<$0.001 & 6.0 [4.6-8.3] & 8.1 [5.9-11.1] & 10.3 [7.5-13.2] & 12.9 [10.1-16.7] & $<$0.001 & 5.6 [4.2-7.4] & 7.3 [5.5-9.9] & 9.7 [6.9-12.9] & 12.5 [9.5-16.5] & $<$0.001 \\
\multicolumn{16}{@{}l}{\textit{Household composition}} \\
\quad \% age 65+ & 18.6 $\pm$ 4.8 & 18.0 $\pm$ 4.9 & 17.5 $\pm$ 4.2 & 16.2 $\pm$ 3.6 & $<$0.001 & 19.5 $\pm$ 4.9 & 18.7 $\pm$ 5.0 & 18.3 $\pm$ 4.2 & 17.0 $\pm$ 3.7 & $<$0.001 & 20.5 $\pm$ 4.9 & 19.6 $\pm$ 4.9 & 19.2 $\pm$ 4.7 & 17.6 $\pm$ 4.2 & $<$0.001 \\
\quad \% age $\leq$17 & 22.2 $\pm$ 3.2 & 21.9 $\pm$ 3.6 & 22.5 $\pm$ 3.1 & 23.8 $\pm$ 3.7 & $<$0.001 & 21.8 $\pm$ 3.2 & 21.9 $\pm$ 3.7 & 22.2 $\pm$ 3.0 & 23.5 $\pm$ 3.7 & $<$0.001 & 21.6 $\pm$ 3.3 & 21.5 $\pm$ 3.3 & 21.9 $\pm$ 3.4 & 23.3 $\pm$ 3.8 & $<$0.001 \\
\quad \% with a disability & 12.8 $\pm$ 2.9 & 15.0 $\pm$ 3.7 & 17.1 $\pm$ 4.3 & 18.4 $\pm$ 4.4 & $<$0.001 & 13.0 $\pm$ 3.0 & 15.1 $\pm$ 3.7 & 17.2 $\pm$ 4.2 & 18.4 $\pm$ 4.5 & $<$0.001 & 13.4 $\pm$ 3.4 & 15.5 $\pm$ 4.0 & 17.3 $\pm$ 4.6 & 17.7 $\pm$ 4.4 & $<$0.001 \\
\quad \% single-parent households & 6.6 $\pm$ 1.8 & 7.7 $\pm$ 1.9 & 8.8 $\pm$ 2.0 & 11.0 $\pm$ 2.8 & $<$0.001 & 6.4 $\pm$ 1.8 & 7.5 $\pm$ 1.9 & 8.6 $\pm$ 2.1 & 10.8 $\pm$ 2.9 & $<$0.001 & 4.3 $\pm$ 1.5 & 5.1 $\pm$ 1.6 & 6.0 $\pm$ 1.9 & 8.1 $\pm$ 2.5 & $<$0.001 \\
\multicolumn{16}{@{}l}{\textit{Minority status \& language}} \\
\quad \% minority & 7.1 [4.7-12.5] & 11.7 [6.1-21.5] & 19.1 [8.9-33.8] & 41.2 [28.1-57.8] & $<$0.001 & 7.5 [4.9-12.4] & 13.1 [7.0-22.7] & 19.6 [9.2-33.2] & 41.8 [28.9-58.2] & $<$0.001 & 7.9 [5.5-12.4] & 12.6 [7.5-21.9] & 20.8 [11.3-33.7] & 44.2 [33.0-59.5] & $<$0.001 \\
\quad \% limited English speaking & 0.40 [0.20-0.90] & 0.70 [0.30-1.7] & 0.90 [0.30-2.4] & 1.3 [0.50-3.5] & $<$0.001 & 0.50 [0.20-0.90] & 0.70 [0.30-1.7] & 0.90 [0.30-2.3] & 1.2 [0.40-3.4] & $<$0.001 & 0.40 [0.20-0.80] & 0.60 [0.20-1.3] & 0.90 [0.30-2.2] & 1.5 [0.60-3.8] & $<$0.001 \\
\multicolumn{16}{@{}l}{\textit{Housing \& transportation}} \\
\quad \% multi-unit housing & 3.3 [1.6-6.0] & 3.5 [1.8-6.8] & 2.6 [1.2-5.8] & 2.0 [1.1-4.3] & $<$0.001 & 3.3 [1.5-6.4] & 3.4 [1.6-7.1] & 2.7 [1.3-6.1] & 2.1 [1.1-4.3] & $<$0.001 & 2.9 [1.4-5.5] & 3.3 [1.5-6.6] & 2.7 [1.2-6.1] & 2.5 [1.3-5.7] & 0.022 \\
\quad \% mobile homes & 5.6 [3.2-9.2] & 9.3 [5.3-14.2] & 13.9 [7.6-20.7] & 19.6 [12.2-27.1] & $<$0.001 & 5.7 [3.2-9.6] & 8.8 [5.1-14.2] & 13.6 [7.3-20.4] & 19.5 [12.0-27.3] & $<$0.001 & 5.5 [3.2-9.2] & 9.3 [5.0-14.8] & 13.6 [7.6-20.6] & 17.4 [9.5-24.9] & $<$0.001 \\
\quad \% crowded housing & 1.2 [0.80-1.6] & 1.7 [1.2-2.4] & 2.2 [1.6-3.0] & 2.9 [2.1-4.2] & $<$0.001 & 1.2 [0.80-1.7] & 1.7 [1.2-2.4] & 2.2 [1.5-3.0] & 2.9 [2.1-4.4] & $<$0.001 & 1.1 [0.70-1.6] & 1.6 [1.2-2.3] & 2.1 [1.4-2.9] & 2.9 [2.0-4.3] & $<$0.001 \\
\quad \% no vehicle & 4.1 [3.1-5.2] & 5.5 [4.3-6.8] & 6.4 [5.0-7.9] & 7.9 [6.4-10.2] & $<$0.001 & 4.0 [3.0-5.1] & 5.3 [4.1-6.9] & 6.1 [4.8-7.8] & 7.8 [6.1-9.9] & $<$0.001 & 4.0 [2.9-5.1] & 5.2 [3.8-6.8] & 5.9 [4.6-7.6] & 7.3 [5.9-9.6] & $<$0.001 \\
\quad \% in group quarters & 1.6 [1.0-2.7] & 1.9 [1.2-4.0] & 2.0 [1.3-3.7] & 2.6 [1.4-5.8] & $<$0.001 & 1.6 [1.0-2.6] & 1.9 [1.2-3.9] & 2.0 [1.3-3.7] & 2.6 [1.5-5.8] & $<$0.001 & 1.6 [1.0-2.7] & 1.8 [1.2-3.7] & 2.0 [1.3-3.7] & 2.6 [1.5-5.6] & $<$0.001 \\
\addlinespace[2pt]
\multicolumn{16}{@{}l}{\textbf{C. Geographic}} \\
\addlinespace[1pt]
\textit{Urban/Rural classification} &  &  &  &  & $<$0.001 &  &  &  &  & $<$0.001 &  &  &  &  & $<$0.001 \\
\quad\quad Large central metro & 4 (0.5) & 10 (1.3) & 27 (3.5) & 22 (2.8) & - & 2 (0.3) & 14 (1.8) & 27 (3.5) & 20 (2.6) & - & 1 (0.1) & 9 (1.1) & 20 (2.6) & 33 (4.2) & - \\
\quad\quad Large fringe metro & 125 (16.0) & 71 (9.1) & 40 (5.1) & 12 (1.5) & - & 129 (16.5) & 68 (8.7) & 38 (4.9) & 13 (1.7) & - & 114 (14.6) & 75 (9.6) & 42 (5.4) & 17 (2.2) & - \\
\quad\quad Medium metro & 64 (8.2) & 100 (12.8) & 101 (12.9) & 55 (7.0) & - & 69 (8.8) & 93 (11.9) & 105 (13.4) & 53 (6.8) & - & 59 (7.5) & 92 (11.7) & 104 (13.3) & 66 (8.4) & - \\
\quad\quad Small metro & 41 (5.2) & 60 (7.7) & 56 (7.2) & 45 (5.7) & - & 38 (4.9) & 67 (8.6) & 52 (6.6) & 45 (5.8) & - & 34 (4.3) & 57 (7.3) & 53 (6.8) & 58 (7.4) & - \\
\quad\quad Micropolitan & 68 (8.7) & 127 (16.2) & 116 (14.8) & 172 (22.0) & - & 68 (8.7) & 116 (14.8) & 129 (16.5) & 170 (21.7) & - & 70 (8.9) & 110 (14.0) & 127 (16.3) & 176 (22.5) & - \\
\quad\quad Noncore & 481 (61.4) & 414 (52.9) & 442 (56.5) & 477 (60.9) & - & 477 (60.9) & 424 (54.2) & 431 (55.1) & 481 (61.5) & - & 505 (64.5) & 440 (56.2) & 435 (55.7) & 433 (55.3) & - \\
\textit{Census division} &  &  &  &  & $<$0.001 &  &  &  &  & $<$0.001 &  &  &  &  & $<$0.001 \\
\quad\quad New England & 39 (5.0) & 18 (2.3) & 8 (1.0) & 2 (0.3) & - & 41 (5.2) & 14 (1.8) & 10 (1.3) & 2 (0.3) & - & 29 (3.7) & 26 (3.3) & 9 (1.2) & 3 (0.4) & - \\
\quad\quad Middle Atlantic & 35 (4.5) & 68 (8.7) & 38 (4.9) & 9 (1.1) & - & 36 (4.6) & 63 (8.1) & 42 (5.4) & 9 (1.2) & - & 37 (4.7) & 60 (7.7) & 40 (5.1) & 13 (1.7) & - \\
\quad\quad East North Central & 153 (19.5) & 157 (20.1) & 108 (13.8) & 19 (2.4) & - & 161 (20.6) & 149 (19.1) & 108 (13.8) & 19 (2.4) & - & 182 (23.2) & 149 (19.0) & 95 (12.2) & 11 (1.4) & - \\
\quad\quad West North Central & 340 (43.4) & 163 (20.8) & 78 (10.0) & 36 (4.6) & - & 322 (41.1) & 171 (21.9) & 82 (10.5) & 42 (5.4) & - & 330 (42.1) & 150 (19.2) & 92 (11.8) & 45 (5.7) & - \\
\quad\quad South Atlantic & 73 (9.3) & 134 (17.1) & 141 (18.0) & 240 (30.7) & - & 80 (10.2) & 135 (17.3) & 145 (18.5) & 228 (29.2) & - & 72 (9.2) & 138 (17.6) & 141 (18.1) & 237 (30.3) & - \\
\quad\quad East South Central & 14 (1.8) & 54 (6.9) & 140 (17.9) & 155 (19.8) & - & 14 (1.8) & 61 (7.8) & 141 (18.0) & 147 (18.8) & - & 23 (2.9) & 80 (10.2) & 146 (18.7) & 115 (14.7) & - \\
\quad\quad West South Central & 26 (3.3) & 69 (8.8) & 158 (20.2) & 217 (27.7) & - & 20 (2.6) & 76 (9.7) & 143 (18.3) & 231 (29.5) & - & 16 (2.0) & 73 (9.3) & 138 (17.7) & 243 (31.0) & - \\
\quad\quad Mountain & 88 (11.2) & 78 (10.0) & 53 (6.8) & 58 (7.4) & - & 91 (11.6) & 70 (9.0) & 58 (7.4) & 57 (7.3) & - & 84 (10.7) & 67 (8.6) & 65 (8.3) & 61 (7.8) & - \\
\quad\quad Pacific & 15 (1.9) & 41 (5.2) & 58 (7.4) & 47 (6.0) & - & 18 (2.3) & 43 (5.5) & 53 (6.8) & 47 (6.0) & - & 10 (1.3) & 40 (5.1) & 55 (7.0) & 55 (7.0) & - \\
\addlinespace[2pt]
\multicolumn{16}{@{}l}{\textbf{D. Demographics}} \\
\addlinespace[1pt]
\quad Total population & 17,642 [6,004-59,008] & 32,820 [12,328-81,205] & 29,299 [14,372-80,520] & 24,911 [13,250-54,540] & $<$0.001 & 19,011 [6,525-64,927] & 31,611 [11,566-81,856] & 30,080 [14,109-81,712] & 24,532 [12,969-49,815] & $<$0.001 & 17,260 [6,228-48,838] & 29,462 [11,806-78,995] & 29,255 [14,144-82,056] & 27,223 [13,481-70,516] & $<$0.001 \\
\quad Daytime population ratio & 0.88 [0.79-0.98] & 0.92 [0.84–1.0] & 0.94 [0.86–1.0] & 0.95 [0.88–1.0] & $<$0.001 & 0.84 [0.75–0.94] & 0.88 [0.80–0.96] & 0.89 [0.81–0.97] & 0.90 [0.83–0.97] & $<$0.001 & 0.85 [0.75-0.94] & 0.90 [0.81-0.98] & 0.94 [0.84-1.0] & 0.94 [0.87-1.0] & $<$0.001 \\
\quad Population density (per sq mi) & 31.7 [6.7-110.3] & 53.9 [18.8-143.3] & 51.3 [23.6-132.6] & 41.0 [21.0-84.7] & $<$0.001 & 32.9 [7.5-123.6] & 54.8 [17.5-147.9] & 48.7 [23.3-142.6] & 39.5 [20.2-81.0] & $<$0.001 & 31.4 [7.6-102.0] & 51.8 [18.3-137.6] & 49.8 [22.4-147.5] & 42.1 [19.8-99.2] & $<$0.001 \\
\quad Sex ratio (Male:Female) & 0.99 [0.97-1.0] & 0.99 [0.96-1.0] & 0.98 [0.95-1.0] & 0.97 [0.94-1.0] & $<$0.001 & 0.99 [0.97-1.0] & 0.99 [0.96-1.0] & 0.98 [0.95-1.0] & 0.98 [0.94-1.0] & $<$0.001 & 1.0 [0.98-1.0] & 0.99 [0.97-1.0] & 0.98 [0.96-1.0] & 0.98 [0.94-1.0] & $<$0.001 \\
\addlinespace[2pt]
\multicolumn{16}{@{}l}{\textbf{G. County Typology (USDA ERS)}} \\
\addlinespace[1pt]
\textit{Economic dependence type} &  &  &  &  & $<$0.001 &  &  &  &  & $<$0.001 &  &  &  &  & $<$0.001 \\
\quad\quad Nonspecialized & 257 (32.8) & 291 (37.2) & 336 (43.0) & 345 (44.1) & - & 260 (33.2) & 295 (37.7) & 341 (43.6) & 332 (42.5) & - & 253 (32.3) & 301 (38.4) & 335 (42.9) & 341 (43.6) & - \\
\quad\quad Farming & 220 (28.1) & 91 (11.6) & 64 (8.2) & 69 (8.8) & - & 211 (26.9) & 96 (12.3) & 67 (8.6) & 70 (9.0) & - & 216 (27.6) & 90 (11.5) & 66 (8.5) & 72 (9.2) & - \\
\quad\quad Mining & 35 (4.5) & 53 (6.8) & 76 (9.7) & 57 (7.3) & - & 28 (3.6) & 51 (6.5) & 74 (9.5) & 68 (8.7) & - & 30 (3.8) & 55 (7.0) & 74 (9.5) & 62 (7.9) & - \\
\quad\quad Manufacturing & 107 (13.7) & 129 (16.5) & 124 (15.9) & 141 (18.0) & - & 106 (13.5) & 133 (17.0) & 123 (15.7) & 139 (17.8) & - & 132 (16.9) & 122 (15.6) & 116 (14.9) & 131 (16.7) & - \\
\quad\quad Federal/State Government & 50 (6.4) & 100 (12.8) & 109 (13.9) & 146 (18.6) & - & 50 (6.4) & 101 (12.9) & 109 (13.9) & 145 (18.5) & - & 46 (5.9) & 102 (13.0) & 111 (14.2) & 146 (18.6) & - \\
\quad\quad Recreation & 114 (14.6) & 118 (15.1) & 73 (9.3) & 25 (3.2) & - & 128 (16.3) & 106 (13.6) & 68 (8.7) & 28 (3.6) & - & 106 (13.5) & 113 (14.4) & 79 (10.1) & 31 (4.0) & - \\
\textit{Population loss county} &  &  &  &  & $<$0.001 &  &  &  &  & $<$0.001 &  &  &  &  & $<$0.001 \\
\quad\quad No & 553 (70.6) & 668 (85.4) & 707 (90.4) & 675 (86.2) & - & 567 (72.4) & 664 (84.9) & 701 (89.6) & 670 (85.7) & - & 557 (71.1) & 665 (84.9) & 691 (88.5) & 690 (88.1) & - \\
\quad\quad Yes & 230 (29.4) & 114 (14.6) & 75 (9.6) & 108 (13.8) & - & 216 (27.6) & 118 (15.1) & 81 (10.4) & 112 (14.3) & - & 226 (28.9) & 118 (15.1) & 90 (11.5) & 93 (11.9) & - \\
\textit{Retirement destination county} &  &  &  &  & $<$0.001 &  &  &  &  & $<$0.001 &  &  &  &  & $<$0.001 \\
\quad\quad No & 698 (89.1) & 646 (82.6) & 655 (83.8) & 692 (88.4) & - & 685 (87.5) & 647 (82.7) & 658 (84.1) & 700 (89.5) & - & 702 (89.7) & 643 (82.1) & 653 (83.6) & 692 (88.4) & - \\
\quad\quad Yes & 85 (10.9) & 136 (17.4) & 127 (16.2) & 91 (11.6) & - & 98 (12.5) & 135 (17.3) & 124 (15.9) & 82 (10.5) & - & 81 (10.3) & 140 (17.9) & 128 (16.4) & 91 (11.6) & - \\
\bottomrule
\end{tabular}%
}

\vspace{6pt}
\begin{minipage}{\dimexpr\textheight+\headheight+\headsep+\footskip\relax}
\scriptsize \textit{Note.} Quartiles are formed within each vintage using overall SVI percentile rank. Q1 = lowest social vulnerability; Q4 = highest. Continuous variables: mean $\pm$ SD by default, median [Q1-Q3] for skewed variables. Categorical variables: n (\%) per level. Primary care physicians and Mental health providers are reported per 100,000 population. p-values across quartiles from one-way ANOVA (continuous), Kruskal-Wallis (skewed), or chi-square (categorical) tests. Empty cells indicate the variable is not defined for that vintage. SVI variable definitions changed in the 2020 release: \% below poverty was replaced by \% below 150\% federal poverty line; per capita income was dropped; \% housing cost-burdened was added; and uninsured moved from adjunct into the socioeconomic theme.
\end{minipage}
\end{landscape}

\clearpage

\section{Association Analysis}

\setcounter{table}{0}
\setcounter{figure}{0}
\renewcommand{\thetable}{\thesection.\arabic{table}}
\renewcommand{\thefigure}{\thesection.\arabic{figure}}

\begin{figure}[ht]
\centering
\includegraphics[width=0.6\textwidth]{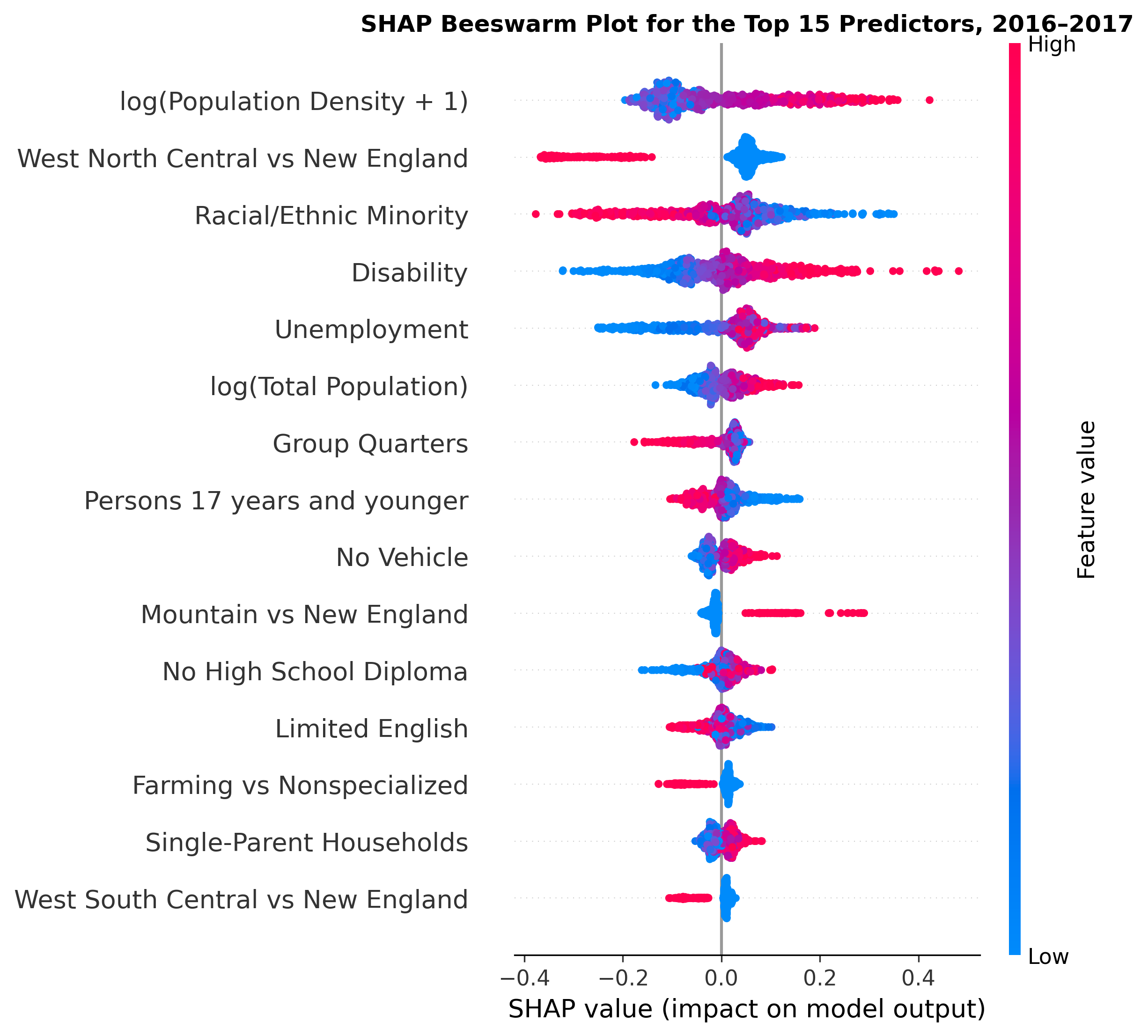}
\caption{SHAP beeswarm plot of top 15 features for XGBoost during 2016-2017.}
\label{fig:beeswarm_top15_2016_2017_XGBoost}
\end{figure}

\begin{figure}[ht]
\centering
\includegraphics[width=0.6\textwidth]{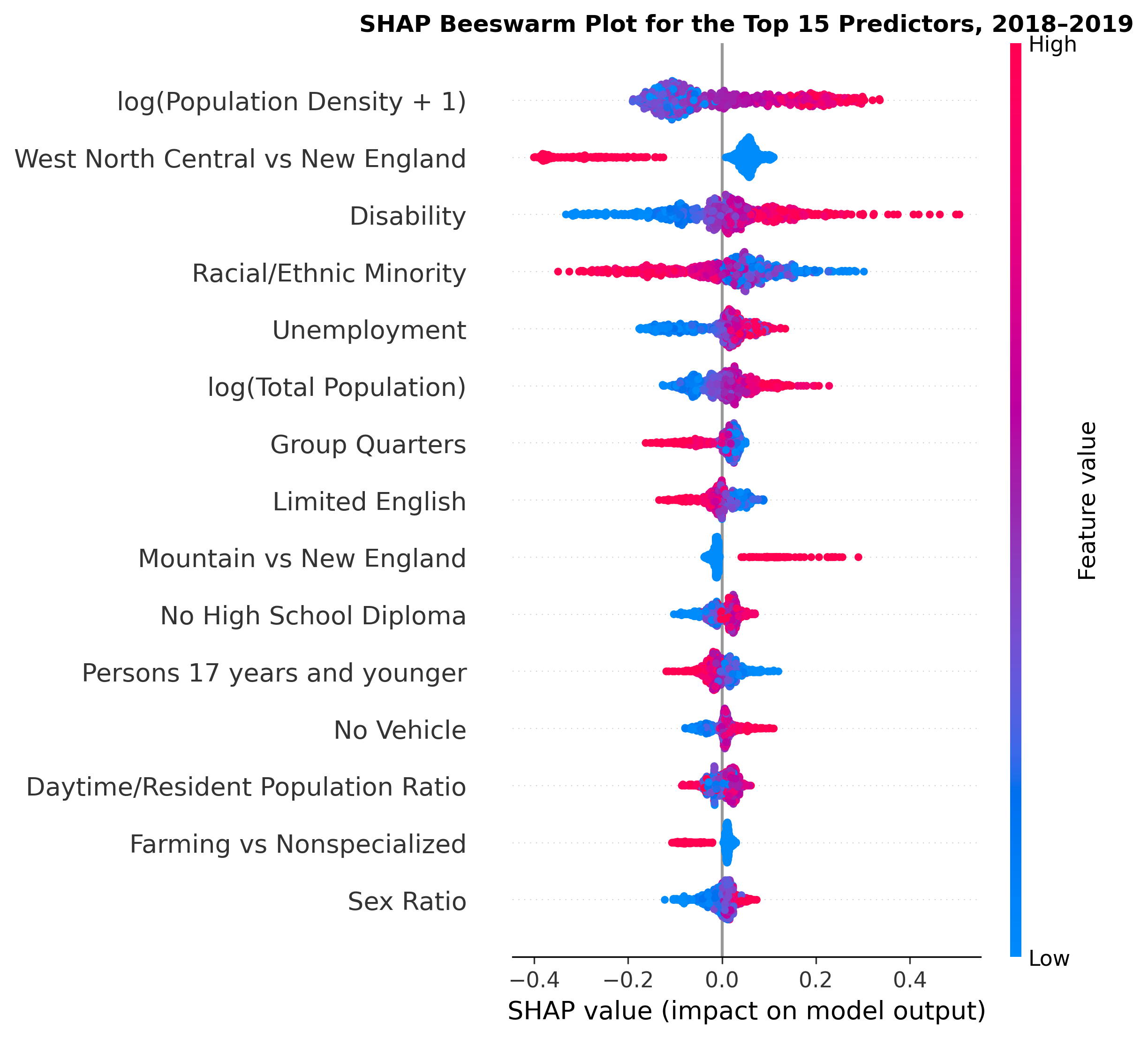}
\caption{SHAP beeswarm plot of top 15 features for XGBoost during 2018-2019.}
\label{fig:beeswarm_top15_2018_2019_XGBoost}
\end{figure}

\begin{table}[htbp]
\centering
\small
\caption{SHAP importance and direction, SVI sub-features. Rows are ordered by mean $|$SHAP$|$ in 2020-2021. Dashes mark features absent from a vintage, following the 2020 SVI revision.}
\label{tab:shap_direction_svi}
\begin{tabular}{lrrrrrr}
\toprule
\multirow{2}{*}{Feature} & \multicolumn{2}{c}{2016-2017} & \multicolumn{2}{c}{2018-2019} & \multicolumn{2}{c}{2020-2021} \\
 & mean $|$SHAP$|$ & $\rho$ & mean $|$SHAP$|$ & $\rho$ & mean $|$SHAP$|$ & $\rho$ \\
\midrule
Racial/Ethnic Minority & 0.0916 & -0.865 & 0.0853 & -0.857 & 0.0868 & -0.897 \\
Disability & 0.0848 & +0.971 & 0.0886 & +0.957 & 0.0825 & +0.950 \\
Unemployment & 0.0719 & +0.772 & 0.0497 & +0.849 & 0.0589 & +0.874 \\
Group Quarters & 0.0349 & -0.764 & 0.0308 & -0.856 & 0.0290 & -0.693 \\
Persons 17 years and younger & 0.0322 & -0.921 & 0.0236 & -0.907 & 0.0217 & -0.788 \\
No High School Diploma & 0.0245 & +0.489 & 0.0238 & +0.664 & 0.0214 & +0.660 \\
Housing Cost Burden & - & - & - & - & 0.0202 & +0.919 \\
Single-Parent Households & 0.0202 & +0.791 & 0.0186 & +0.623 & 0.0197 & +0.499 \\
Limited English & 0.0237 & -0.609 & 0.0275 & -0.876 & 0.0196 & -0.709 \\
Mobile Homes & 0.0119 & +0.151 & 0.0156 & -0.266 & 0.0191 & +0.278 \\
Below 150\% Poverty Level & - & - & - & - & 0.0178 & +0.705 \\
Multi-Unit Housing & 0.0093 & -0.552 & 0.0158 & -0.700 & 0.0118 & -0.794 \\
Persons 65+ years & 0.0169 & -0.860 & 0.0137 & -0.497 & 0.0115 & -0.422 \\
Uninsured & 0.0147 & +0.647 & 0.0103 & -0.356 & 0.0110 & -0.562 \\
Crowded Housing & 0.0095 & +0.168 & 0.0089 & +0.802 & 0.0108 & +0.706 \\
No Vehicle & 0.0270 & +0.909 & 0.0230 & +0.865 & 0.0087 & +0.460 \\
Per Capita Income & 0.0112 & -0.052 & 0.0148 & +0.206 & - & - \\
Below Poverty & 0.0114 & +0.047 & 0.0144 & +0.715 & - & - \\
\bottomrule
\end{tabular}
\end{table}

\begin{table}[htbp]
\centering
\small
\caption{SHAP importance and direction, continuous non-SVI covariates. Rows are ordered by mean $|$SHAP$|$ in 2020-2021. Dashes mark features absent from a vintage, following the 2020 SVI revision.}
\label{tab:shap_direction_cov}
\begin{tabular}{lrrrrrr}
\toprule
\multirow{2}{*}{Feature} & \multicolumn{2}{c}{2016-2017} & \multicolumn{2}{c}{2018-2019} & \multicolumn{2}{c}{2020-2021} \\
 & mean $|$SHAP$|$ & $\rho$ & mean $|$SHAP$|$ & $\rho$ & mean $|$SHAP$|$ & $\rho$ \\
\midrule
log(Population Density + 1) & 0.1130 & +0.869 & 0.1151 & +0.866 & 0.1186 & +0.900 \\
log(Total Population) & 0.0380 & +0.950 & 0.0435 & +0.955 & 0.0300 & +0.928 \\
Sex Ratio & 0.0155 & +0.636 & 0.0209 & +0.789 & 0.0159 & +0.786 \\
Daytime/Resident Population Ratio & 0.0139 & +0.195 & 0.0215 & +0.464 & 0.0132 & +0.412 \\
\bottomrule
\end{tabular}
\end{table}

\begin{table}[htbp]
\centering
\small
\caption{SHAP importance and direction, binary indicators. Rows are ordered by mean $|$SHAP$|$ in 2020-2021.}
\label{tab:shap_direction_ind}
\begin{tabular}{lrrrrrr}
\toprule
\multirow{2}{*}{Feature} & \multicolumn{2}{c}{2016-2017} & \multicolumn{2}{c}{2018-2019} & \multicolumn{2}{c}{2020-2021} \\
 & mean $|$SHAP$|$ & mean SHAP & mean $|$SHAP$|$ & mean SHAP & mean $|$SHAP$|$ & mean SHAP \\
\midrule
West North Central vs New England & 0.0992 & -0.278 & 0.1088 & -0.310 & 0.1101 & -0.315 \\
Mountain vs New England & 0.0269 & +0.133 & 0.0242 & +0.125 & 0.0241 & +0.120 \\
Farming vs Nonspecialized & 0.0233 & -0.075 & 0.0215 & -0.071 & 0.0204 & -0.067 \\
West South Central vs New England & 0.0190 & -0.066 & 0.0208 & -0.068 & 0.0165 & -0.049 \\
East North Central vs New England & 0.0119 & -0.042 & 0.0139 & -0.052 & 0.0110 & -0.043 \\
Large Fringe Metro vs Noncore & 0.0090 & +0.053 & 0.0084 & +0.054 & 0.0090 & +0.054 \\
Mining vs Nonspecialized & 0.0095 & +0.062 & 0.0079 & +0.054 & 0.0082 & +0.054 \\
Recreation vs Nonspecialized & 0.0094 & +0.044 & 0.0105 & +0.048 & 0.0074 & +0.032 \\
East South Central vs New England & 0.0052 & -0.021 & 0.0072 & -0.027 & 0.0070 & -0.020 \\
Manufacturing vs Nonspecialized & 0.0114 & -0.035 & 0.0114 & -0.035 & 0.0065 & -0.020 \\
Middle Atlantic vs New England & 0.0056 & -0.062 & 0.0073 & -0.078 & 0.0054 & -0.057 \\
Federal/State Government vs Nonspecialized & 0.0020 & +0.009 & 0.0002 & +0.000 & 0.0038 & +0.015 \\
Small Metro vs Noncore & 0.0040 & -0.038 & 0.0023 & -0.021 & 0.0030 & -0.024 \\
Population Loss County & 0.0027 & +0.012 & 0.0015 & -0.004 & 0.0028 & +0.008 \\
South Atlantic vs New England & 0.0020 & +0.005 & 0.0043 & +0.011 & 0.0024 & +0.007 \\
Retirement Destination County & 0.0017 & -0.005 & 0.0005 & -0.000 & 0.0023 & -0.007 \\
Micropolitan vs Noncore & 0.0022 & -0.004 & 0.0011 & -0.003 & 0.0021 & -0.005 \\
Pacific vs New England & 0.0018 & +0.015 & 0.0031 & +0.028 & 0.0013 & +0.003 \\
Medium Metro vs Noncore & 0.0023 & +0.011 & 0.0004 & +0.002 & 0.0004 & -0.002 \\
Large Central Metro vs Noncore & 0.0001 & +0.002 & 0.0000 & +0.001 & 0.0000 & +0.000 \\
\bottomrule
\end{tabular}
\end{table}

\begin{table}[htbp]
\centering
\small
\caption{Aggregated SHAP magnitude and direction by SVI theme and for overall SVI.}
\label{tab:theme_shap_magnitude}
\begin{tabular}{lrrrrrr}
\toprule
\multirow{2}{*}{Measure} & \multicolumn{2}{c}{2016-2017} & \multicolumn{2}{c}{2018-2019} & \multicolumn{2}{c}{2020-2021} \\
 & mean $|$SHAP$|$ & $\rho$ & mean $|$SHAP$|$ & $\rho$ & mean $|$SHAP$|$ & $\rho$ \\
\midrule
Socioeconomic & 0.0920 & +0.573 & 0.0758 & +0.772 & 0.0857 & +0.714 \\
Household & 0.0948 & +0.553 & 0.0984 & +0.574 & 0.1022 & +0.266 \\
Minority status & 0.1048 & -0.805 & 0.1020 & -0.854 & 0.0868 & -0.897 \\
Housing/Transportation & 0.0435 & +0.049 & 0.0454 & +0.047 & 0.0415 & -0.052 \\
Overall SVI & 0.2153 & +0.242 & 0.2127 & +0.249 & 0.2033 & +0.163 \\
\bottomrule
\end{tabular}
\end{table}

\begin{figure}[htbp]
    \centering
    \includegraphics[width=0.59\textwidth]{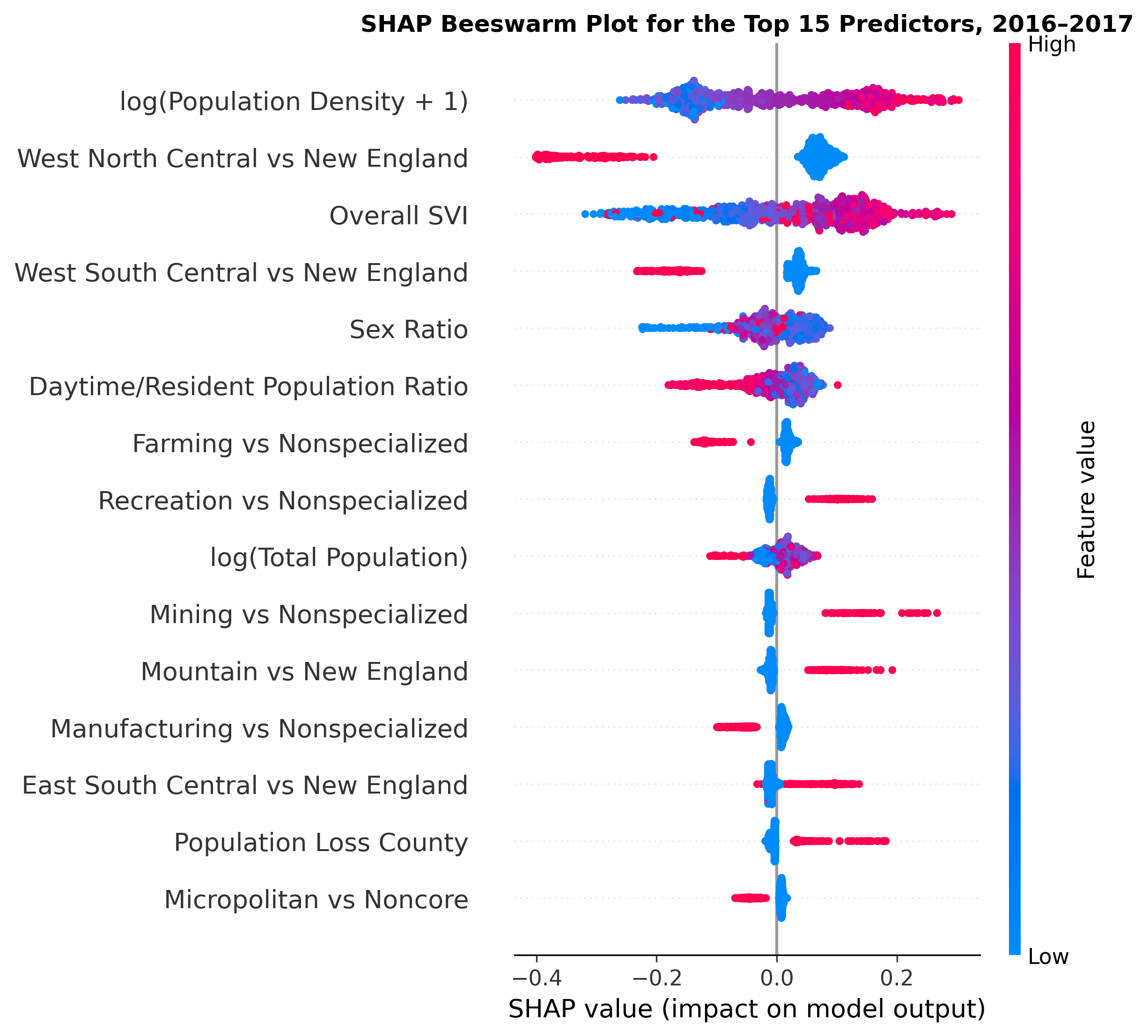}
    \caption{SHAP beeswarm plot for the overall-SVI XGBoost association model, 2016-2017.}
    \label{fig:overall_svi_shap_2016_2017}
\end{figure}

\begin{figure}[htbp]
    \centering
    \includegraphics[width=0.59\textwidth]{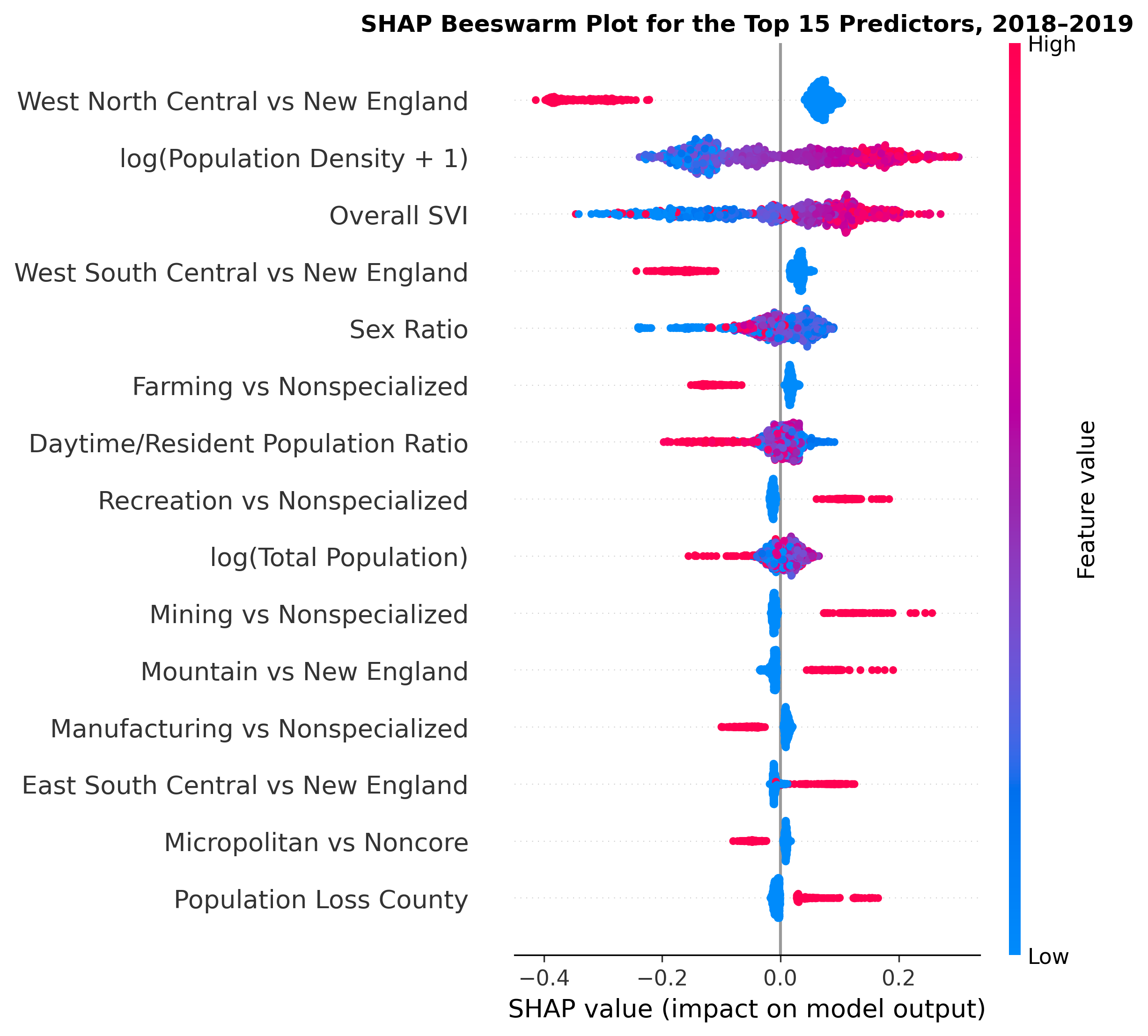}
    \caption{SHAP beeswarm plot for the overall-SVI XGBoost association model, 2018-2019.}
    \label{fig:overall_svi_shap_2018_2019}
\end{figure}

\begin{figure}[htbp]
    \centering
    \includegraphics[width=0.59\textwidth]{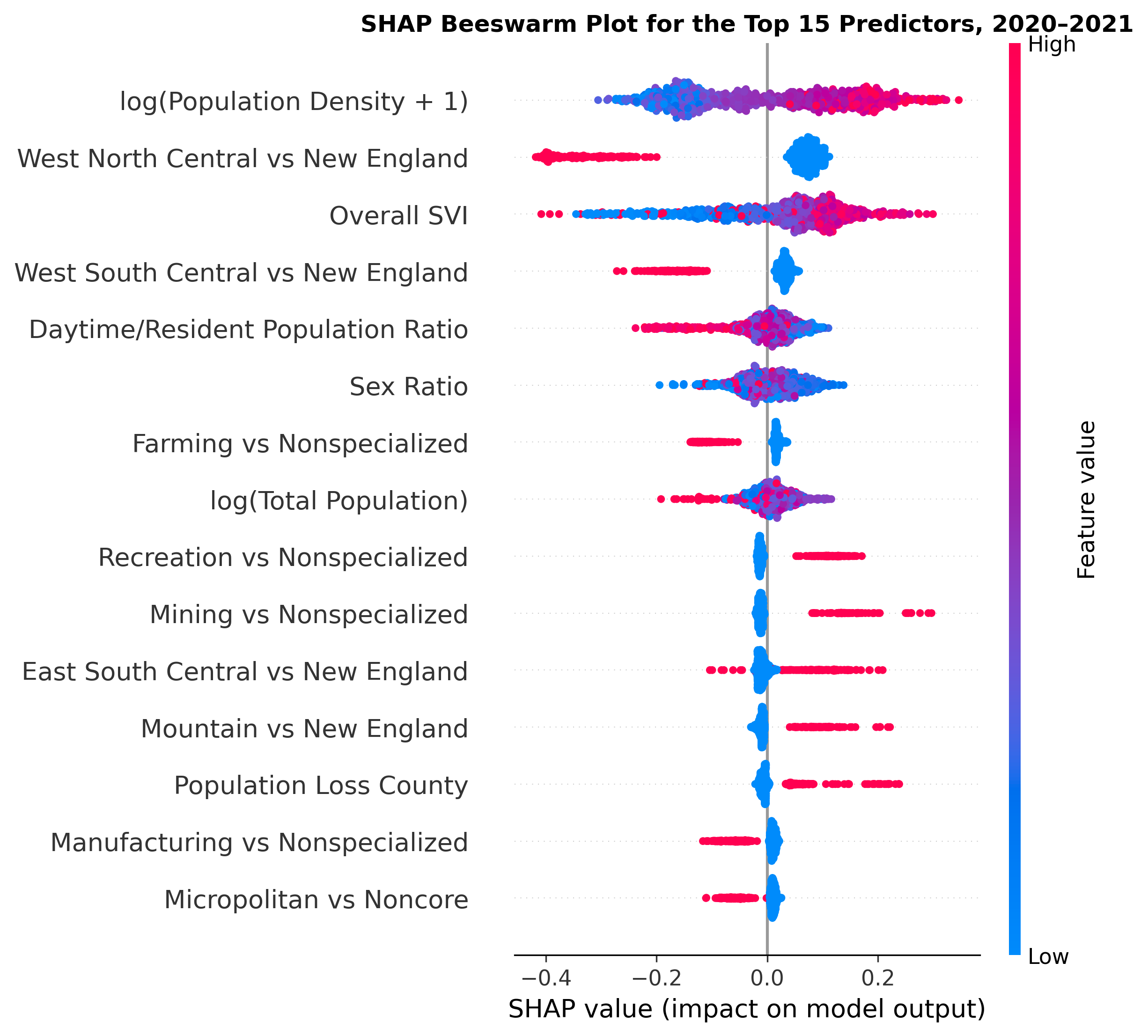}
    \caption{SHAP beeswarm plot for the overall-SVI XGBoost association model, 2020-2021.}
    \label{fig:overall_svi_shap_2020_2021}
\end{figure}

\begin{figure}[ht]
\centering
\includegraphics[width=0.59\textwidth]{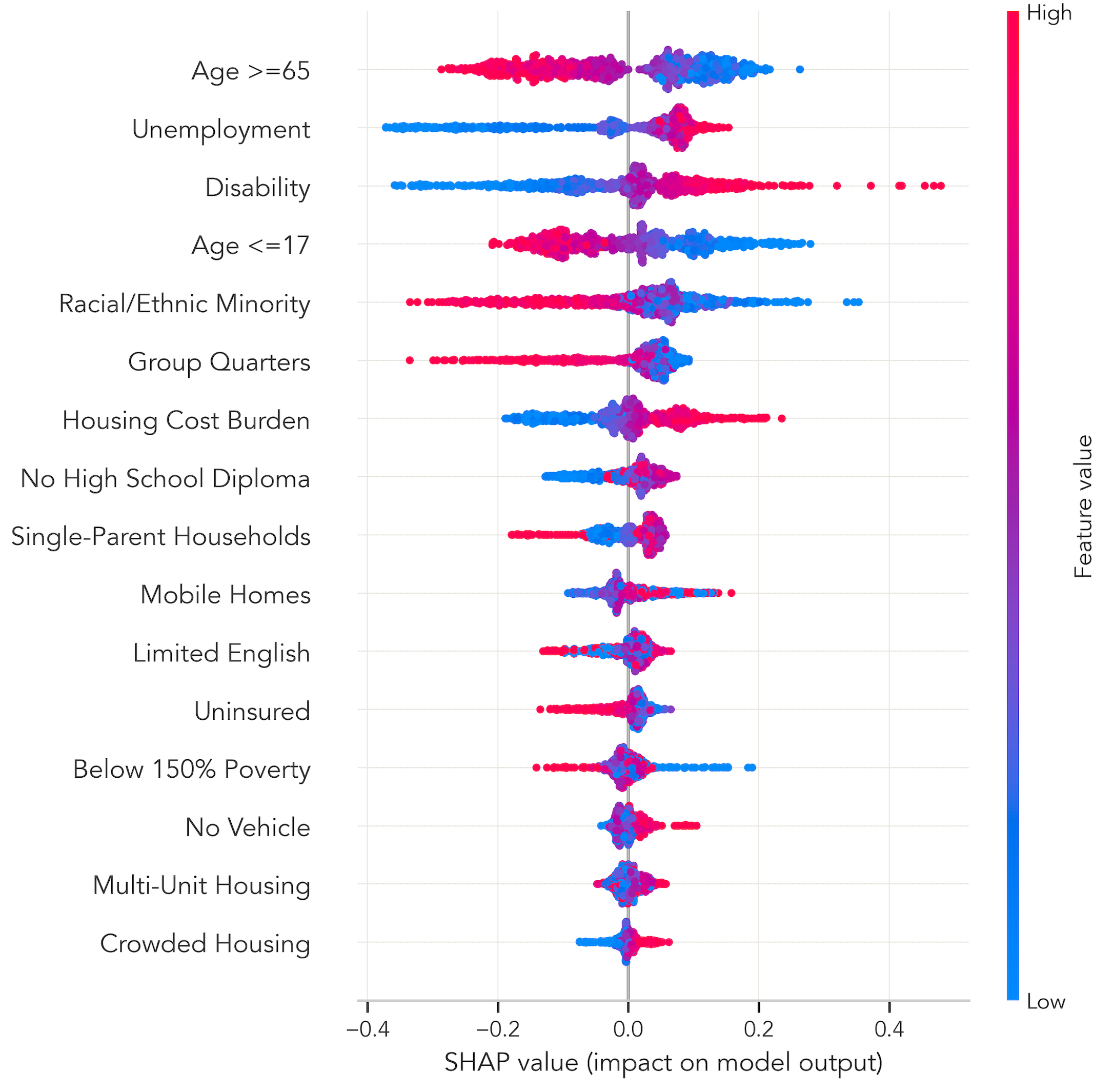}
\caption{SHAP beeswarm plot of SVI features for XGBoost during 2020-2021.}
\label{fig:beeswarm_svi_2020_2021_XGBoost}
\end{figure}

\begin{figure}[ht]
\centering
\includegraphics[width=0.6\textwidth]{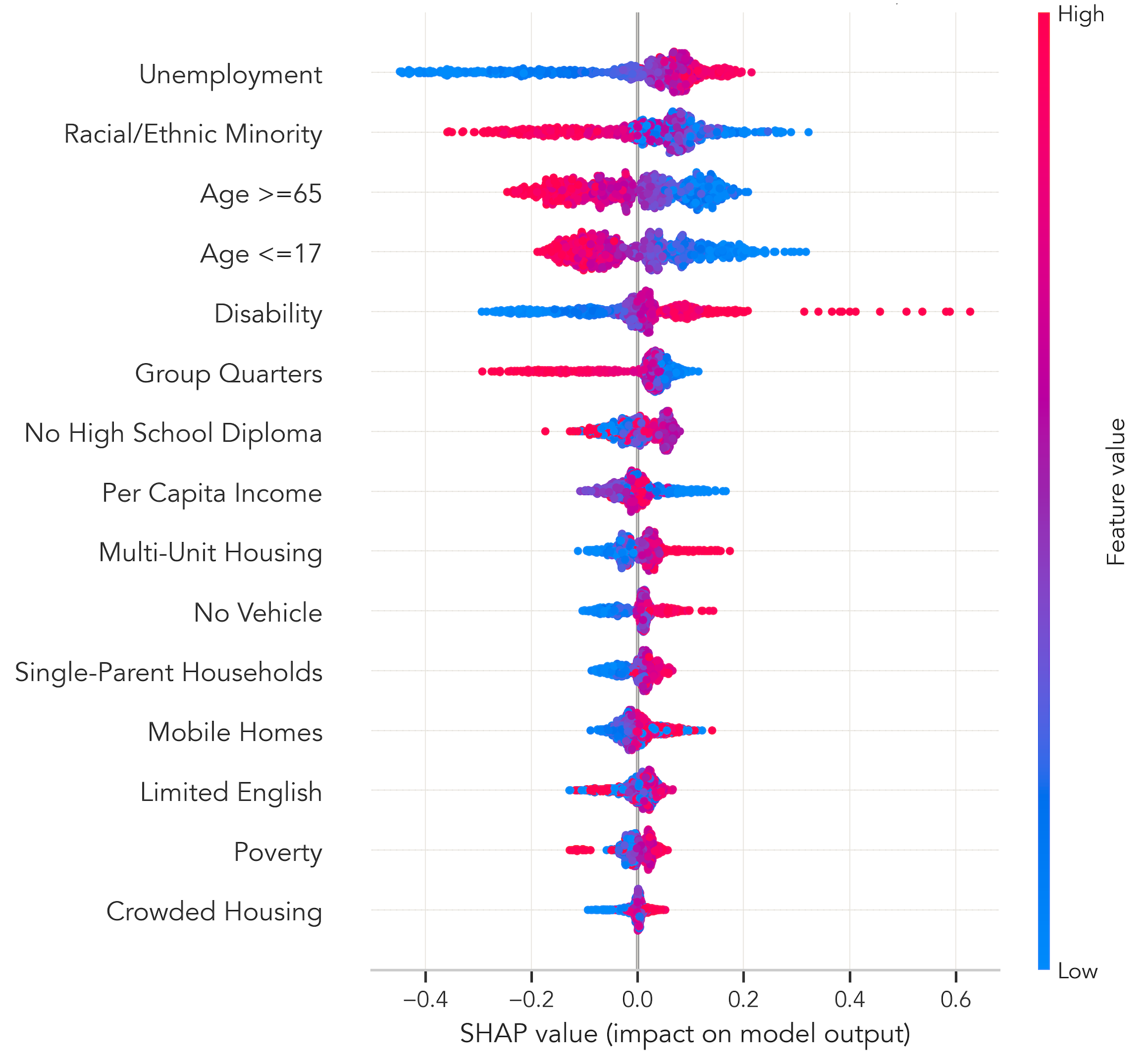}
\caption{SHAP beeswarm plot of SVI features for XGBoost during 2018-2019.}
\label{fig:beeswarm_svi_2018_2019_XGBoost}
\end{figure}

\begin{figure}[ht]
\centering
\includegraphics[width=0.6\textwidth]{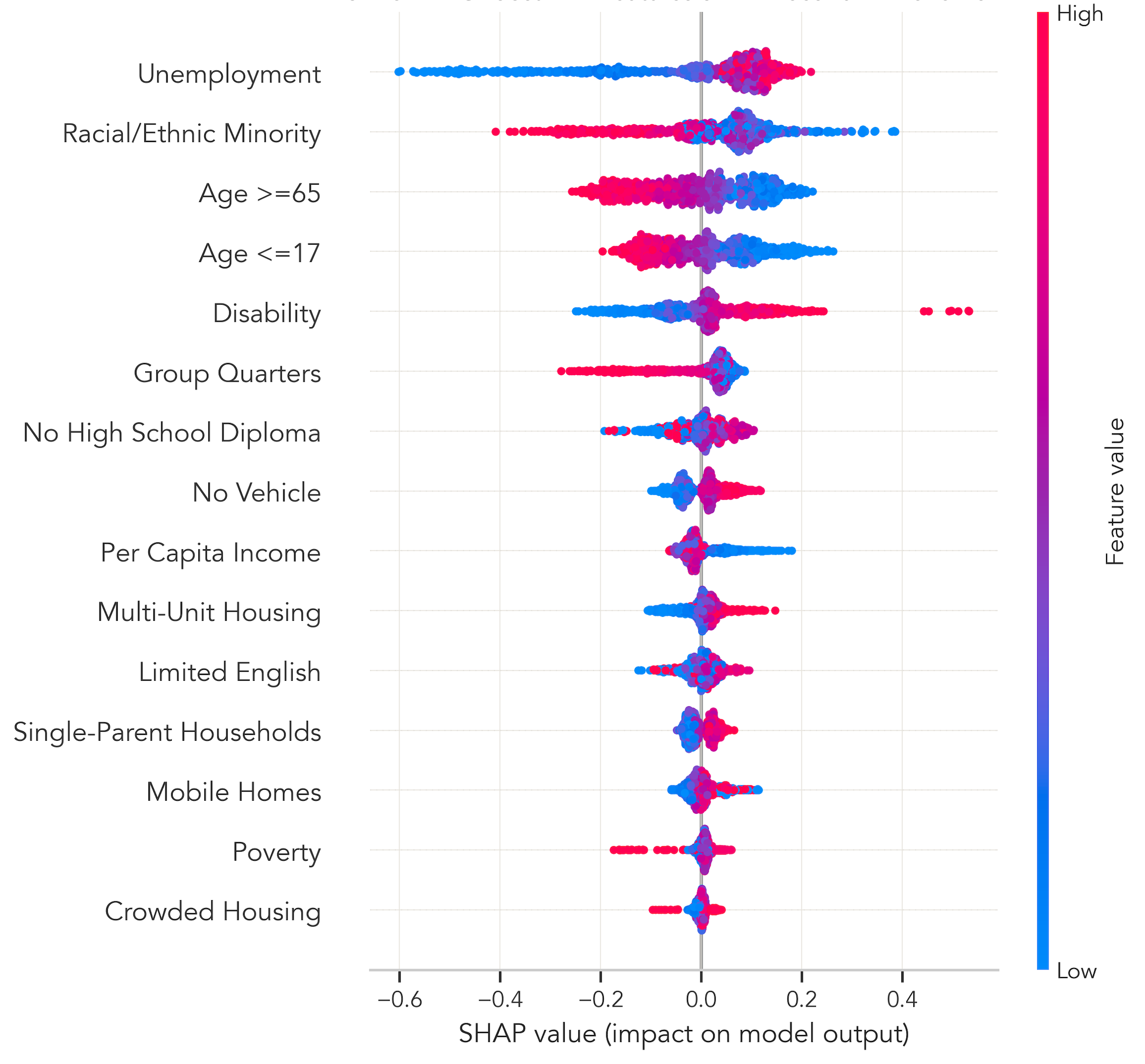}
\caption{SHAP beeswarm plot of SVI features for XGBoost during 2016-2017.}
\label{fig:beeswarm_svi_2016_2017_XGBoost}
\end{figure}

\begin{figure}[ht]
\centering
\includegraphics[width=0.6\textwidth]{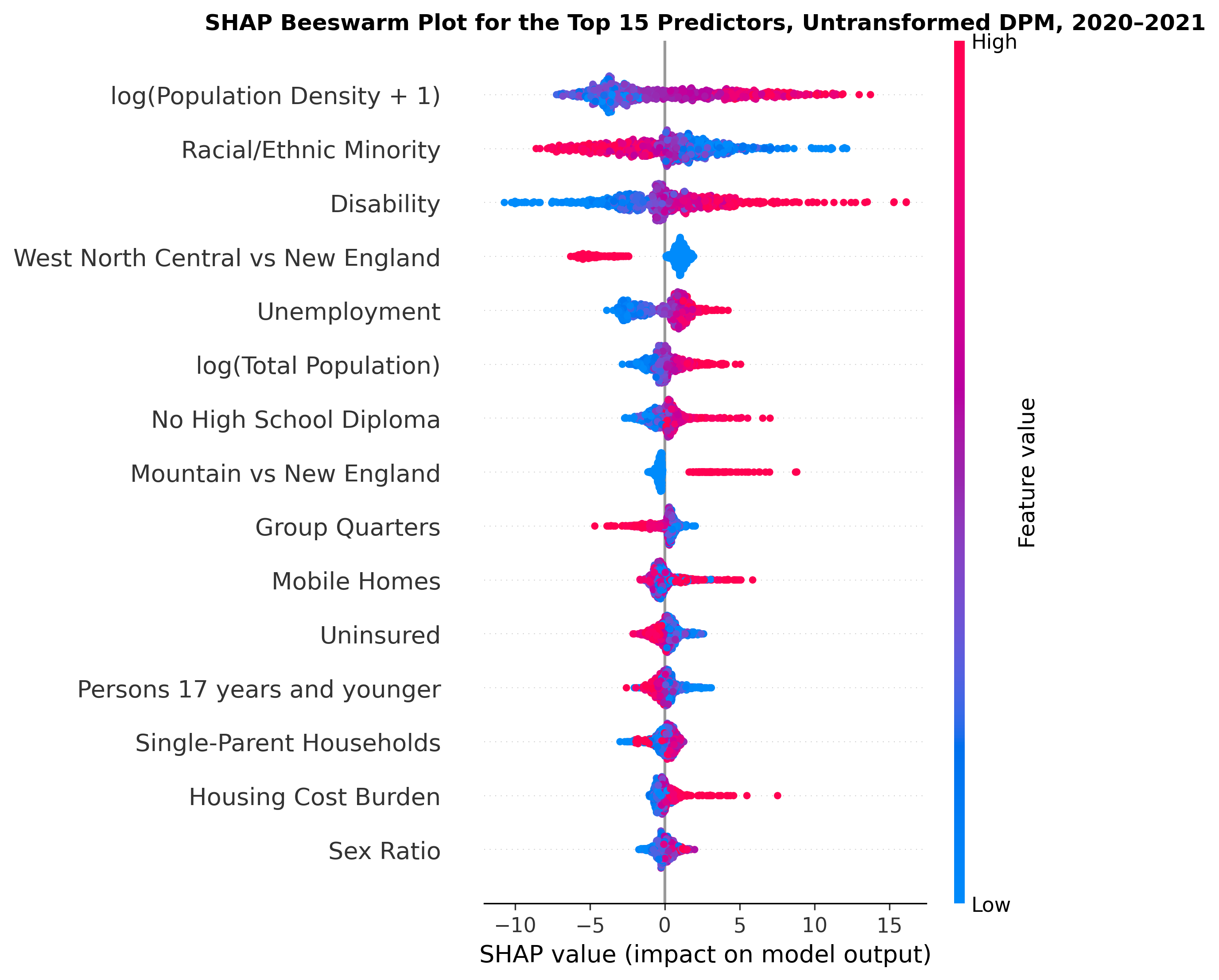}
\caption{SHAP beeswarm plot using untransformed drug poisoning mortality as outcome for XGBoost during 2020-2021.}
\label{fig:rawdpm_beeswarm_svi_2020_2021_XGBoost}
\end{figure}

\begin{figure}[ht]
\centering
\includegraphics[width=0.6\textwidth]{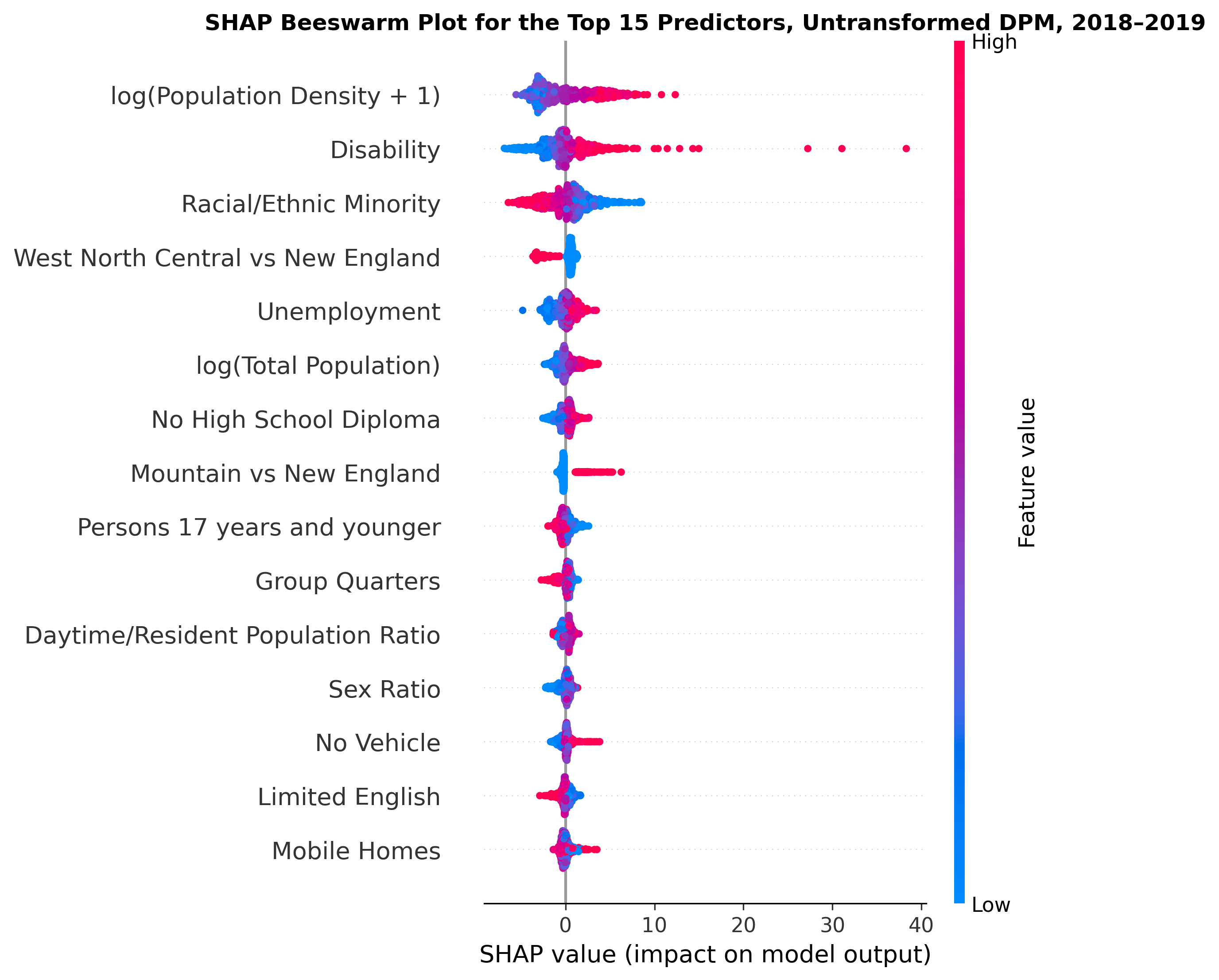}
\caption{SHAP beeswarm plot using untransformed drug poisoning mortality as outcome for XGBoost during 2018-2019.}
\label{fig:rawdpm_beeswarm_svi_2018_2019_XGBoost}
\end{figure}

\begin{figure}[ht]
\centering
\includegraphics[width=0.6\textwidth]{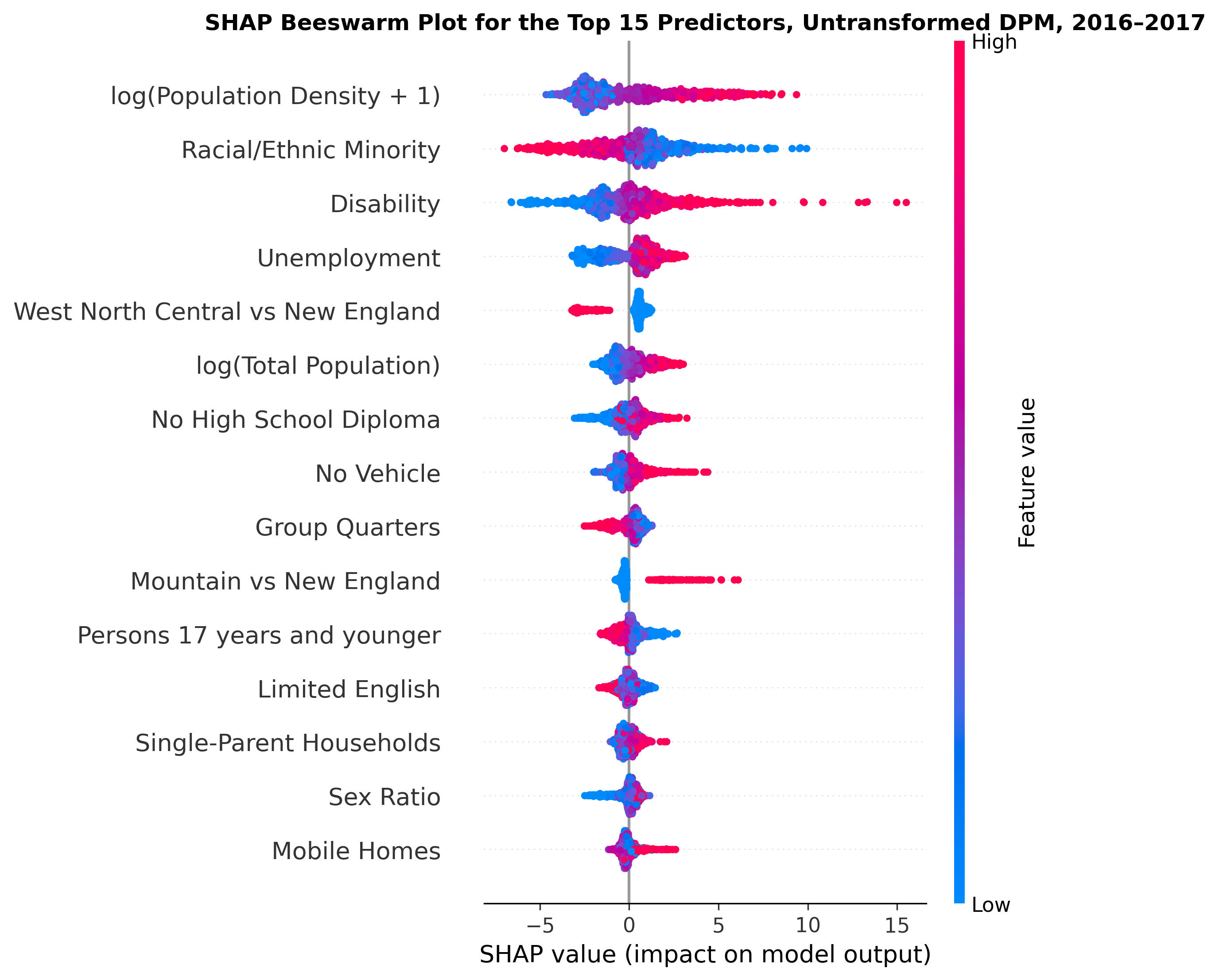}
\caption{SHAP beeswarm plot using untransformed drug poisoning mortality as outcome for XGBoost during 2016-2017.}
\label{fig:rawdpm_beeswarm_svi_2016_2017_XGBoost}
\end{figure}

\clearpage

\section{Double Machine Learning Analysis}

\setcounter{table}{0}
\setcounter{figure}{0}
\renewcommand{\thetable}{\thesection.\arabic{table}}
\renewcommand{\thefigure}{\thesection.\arabic{figure}}

\begin{figure}[htbp]
\centering
\includegraphics[width=0.6\textwidth]{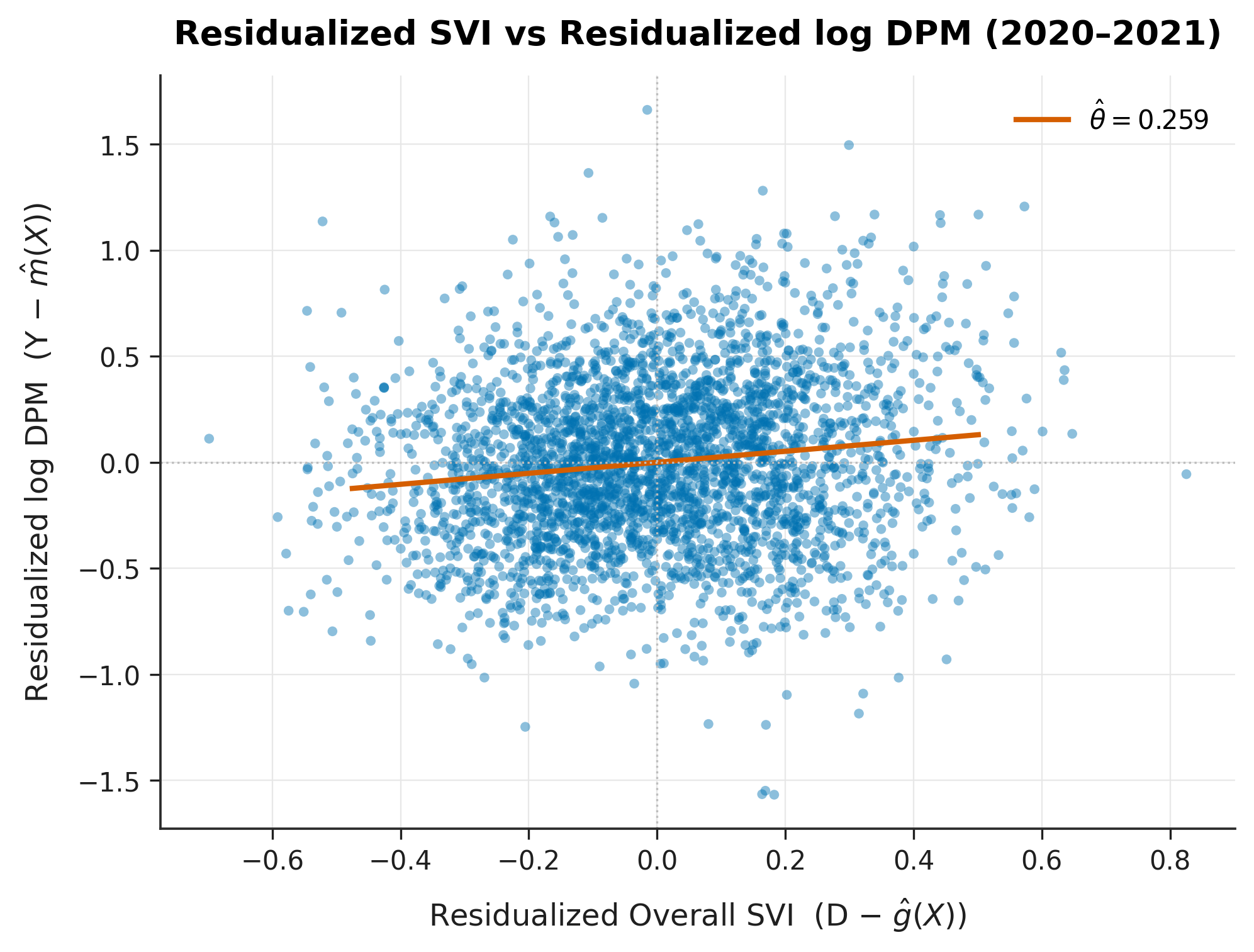}
\caption{Partialling-out plot for the linear DML benchmark specification, 2020-2021. Points show residualized log DPM $(Y-\hat{m}(X))$ plotted against residualized Overall SVI $(D-\hat{g}(X))$. The fitted line represents the final-stage DML estimate.}
\label{fig:dml_residual_scatter_2020_2021}
\end{figure}

\begin{figure}[htbp]
\centering
\includegraphics[width=\textwidth]{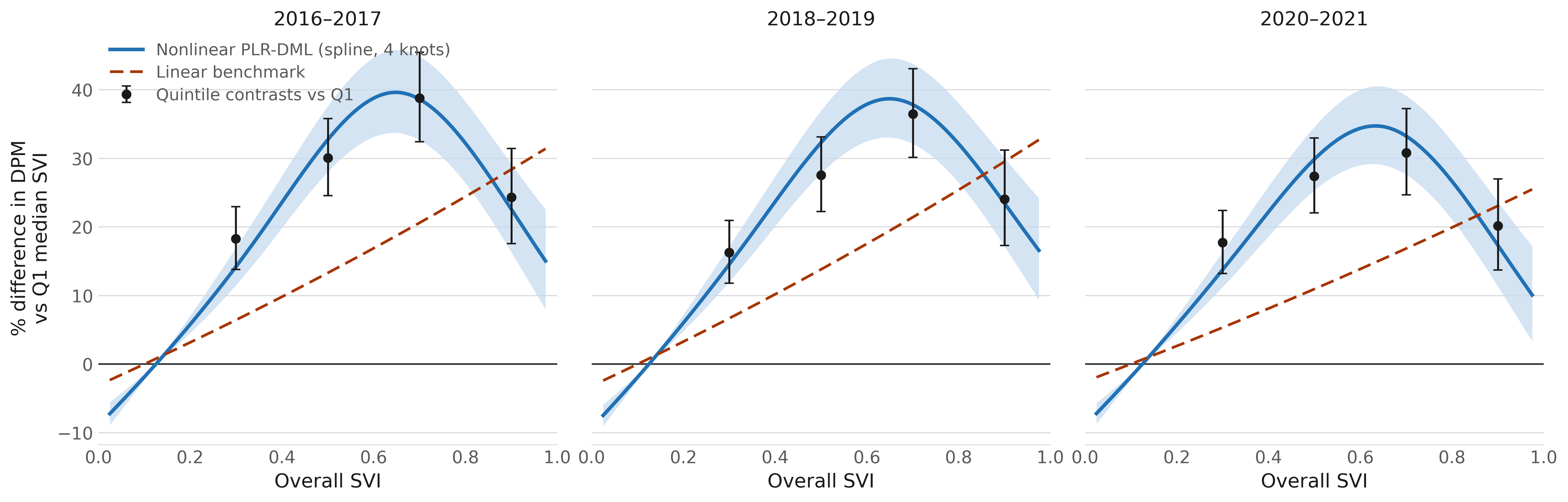}
\caption{Adjusted SVI-mortality relationship with quintile contrasts overlaid.}
\label{fig:dml_nonlinear_quintile}
\end{figure}

\begin{table}[htbp]
\centering
\small
\setlength{\tabcolsep}{5pt}
\caption{Nonlinear fit against the linear benchmark, evaluated at the same values of overall SVI. Entries are percent differences in drug poisoning mortality relative to a county at the first-quartile median SVI. T}
\label{tab:decile_nonlinear_vs_linear}
\begin{tabular}{lrrrrrr}
\toprule
\multirow{2}{*}{Overall SVI} & \multicolumn{2}{c}{2016-2017} & \multicolumn{2}{c}{2018-2019} & \multicolumn{2}{c}{2020-2021} \\
 & nonlinear & linear & nonlinear & linear & nonlinear & linear \\
\midrule
0.10 & -1.9 & -0.8 & -1.9 & -0.8 & -1.9 & -0.6 \\
0.20 & +5.8 & +2.4 & +6.0 & +2.5 & +5.7 & +2.0 \\
0.30 & +14.2 & +5.6 & +14.6 & +5.8 & +13.8 & +4.6 \\
0.40 & +23.6 & +9.0 & +23.7 & +9.3 & +22.2 & +7.4 \\
0.50 & +32.7 & +12.4 & +32.3 & +12.9 & +29.9 & +10.2 \\
0.60 & +38.8 & +16.0 & +37.9 & +16.6 & +34.4 & +13.1 \\
0.70 & +38.6 & +19.7 & +37.8 & +20.5 & +33.2 & +16.1 \\
0.80 & +32.2 & +23.5 & +32.1 & +24.4 & +26.7 & +19.1 \\
0.90 & +22.6 & +27.4 & +23.4 & +28.5 & +17.3 & +22.3 \\
0.95 & +17.5 & +29.4 & +18.8 & +30.6 & +12.4 & +23.9 \\
\bottomrule
\end{tabular}
\end{table}

\begin{table}[htbp]
\centering
\begin{threeparttable}
\caption{Sensitivity analyses for the nonlinear specification.}
\label{tab:sensitivity_nonlinear}
\renewcommand{\arraystretch}{1.2}
\small
\begin{tabular}{lrlcrlr}
\hline
& \multicolumn{2}{c}{Linearity} &
\begin{tabular}[c]{@{}c@{}}Peak $d^*$\\ 3/4/5 knots\end{tabular} &
\begin{tabular}[c]{@{}c@{}}0.25–0.75\\ contrast\end{tabular} &
\begin{tabular}[c]{@{}c@{}}Q5 vs Q4\\ (\%, $p$)\end{tabular} &
\begin{tabular}[c]{@{}c@{}}Linear\\ per 0.1\end{tabular} \\
\cline{2-3}
Period & Wald & $p$ & & & & \\
\hline
\multicolumn{7}{l}{\textit{Primary (XGBoost, census division)}} \\
\quad 2016-2017 & 130.0 & $5.92 \times 10^{-29}$ & 0.60 / 0.65 / 0.69 & 23.8\% & $-$10.4\% ($<0.001$) & 3.17\% \\
\quad 2018-2019 & 123.6 & $1.41 \times 10^{-27}$ & 0.61 / 0.65 / 0.68 & 23.0\% & $-$9.1\% ($<0.001$) & 3.29\% \\
\quad 2020-2021 & 130.1 & $5.59\times10^{-29}$ & 0.58 / 0.63 / 0.66 & 18.9\% & $-$8.1\% ($<0.001$) & 2.63\% \\
\addlinespace
\multicolumn{7}{l}{\textit{State-indicator adjustment}} \\
\quad 2016-2017 & 68.2 & $1.56\times10^{-15}$ & 0.69 / 0.68 / 0.71 & 20.6\% & $-$4.8\% (0.024) & 3.25\% \\
\quad 2018-2019 & 74.3 & $7.22\times10^{-17}$ & 0.70 / 0.68 / 0.72 & 20.9\% & $-$5.6\% (0.008) & 3.35\% \\
\quad 2020-2021 & 76.5 & $2.41\times10^{-17}$ & 0.65 / 0.67 / 0.71 & 18.8\% & $-$6.1\% (0.003) & 2.83\% \\
\addlinespace
\multicolumn{7}{l}{\textit{Elastic net nuisance}} \\
\quad 2016-2017 & 212.0 & $9.38\times10^{-47}$ & 0.57 / 0.64 / 0.70 & 27.7\% & $-$15.2\% ($<0.001$) & 2.99\% \\
\quad 2018-2019 & 177.2 & $3.35\times10^{-39}$ & 0.58 / 0.64 / 0.70 & 25.7\% & $-$13.8\% ($<0.001$) & 2.97\% \\
\quad 2020-2021 & 202.4 & $1.15\times10^{-44}$ & 0.55 / 0.62 / 0.68 & 18.6\% & $-$13.5\% ($<0.001$) & 2.06\% \\
\addlinespace
\multicolumn{7}{l}{\textit{Random Forest nuisance}} \\
\quad 2016-2017 & 153.2 & $5.41\times10^{-34}$ & 0.59 / 0.63 / 0.68 & 19.6\% & $-$12.1\% ($<0.001$) & 3.11\% \\
\quad 2018-2019 & 116.4 & $5.35\times10^{-26}$ & 0.61 / 0.64 / 0.69 & 20.6\% & $-$9.9\% ($<0.001$) & 3.27\% \\
\quad 2020-2021 & 127.0 & $2.61\times10^{-28}$ & 0.58 / 0.63 / 0.69 & 18.4\% & $-$9.4\% ($<0.001$) & 2.48\% \\
\hline
\end{tabular}
\begin{tablenotes}[flushleft]
\small
\item \textit{Note.} Wald tests assess nonlinearity under the four-knot specification. Peak SVI values are shown for three-, four-, and five-knot splines. The IQR contrast compares SVI values of 0.25 and 0.75. Q5-Q4 summarizes the upper-tail contrast.
\end{tablenotes}
\end{threeparttable}
\end{table}

\begin{table}[htbp]
\centering
\begin{threeparttable}
\caption{Outcome-scale sensitivity of the nonlinear specification.}
\label{tab:sensitivity_rawdpm_nonlinear}
\renewcommand{\arraystretch}{1.2}
\small
\begin{tabular}{lrcccl}
\hline
Period &
\begin{tabular}[c]{@{}c@{}}Wald\\(4 knots)\end{tabular} &
\begin{tabular}[c]{@{}c@{}}Linearity $p$\\3 / 4 / 5 knots\end{tabular} &
\begin{tabular}[c]{@{}c@{}}Peak\\SVI\end{tabular} &
\begin{tabular}[c]{@{}c@{}}Peak bin\\quart. / quint.\end{tabular} &
\begin{tabular}[c]{@{}c@{}}Q5 vs Q4\\$p$\end{tabular} \\
\hline
2016-2017 & 65.0 &
$2.73{\times}10^{-12}$ / $7.85{\times}10^{-15}$ / $1.98{\times}10^{-16}$ &
0.68 & Q3 / Q4 & 0.012 \\

2018-2019 & 63.3 &
$5.45{\times}10^{-13}$ / $1.81{\times}10^{-14}$ / $8.76{\times}10^{-16}$ &
0.68 & Q4 / Q4 & 0.012 \\

2020-2021 & 87.8 &
$2.79{\times}10^{-16}$ / $8.48{\times}10^{-20}$ / $2.03{\times}10^{-20}$ &
0.65 & Q3 / Q4 & 0.011 \\
\hline
\end{tabular}
\begin{tablenotes}[flushleft]
\small
\item \textit{Note.} The nonlinear specification was re-estimated using
untransformed drug poisoning mortality, with all other specifications
unchanged. Wald statistics are reported for the four-knot specification.
Linearity $p$ values are shown in the order of three-, four-, and five-knot
splines, tested on one, two, and three degrees of freedom, respectively.
The Q5-versus-Q4 contrast is negative in every window. On the log scale,
the fitted maxima are 0.65, 0.65, and 0.63.
\end{tablenotes}
\end{threeparttable}
\end{table}

\begin{table}[!t]
\centering
\begin{threeparttable}
\caption{Linear PLR-DML benchmark estimates using untransformed drug poisoning mortality.}
\label{tab:sensitivity_rawdpm_linear}
\renewcommand{\arraystretch}{1.25}
\small
\begin{tabular}{lllllc}
\hline
Period & $\hat{\theta}$ & SE & 95\% CI & p-value &
\begin{tabular}[c]{@{}c@{}}Per 0.10\\ SVI\end{tabular} \\
\hline
2016-2017 & 6.460 & 0.705 & [5.078, 7.841] & $<0.001$ & 0.646 \\
2018-2019 & 6.893 & 0.682 & [5.556, 8.229] & $<0.001$ & 0.689 \\
2020-2021 & 7.888 & 1.031 & [5.868, 9.908] & $<0.001$ & 0.789 \\
\hline
\end{tabular}
\begin{tablenotes}[flushleft]
\small
\item \textit{Note.} The linear benchmark re-estimated was re-estimated using untransformed DPM with all other specifications unchanged. $\hat{\theta}$ is in deaths per 100,000 per unit of overall SVI.
\end{tablenotes}
\end{threeparttable}
\end{table}